\documentclass[twocolumn]{aastex61}

\pdfoutput=1 
\usepackage{amsmath,amstext}
\usepackage[T1]{fontenc}
\usepackage{apjfonts} 
\usepackage[figure,figure*]{hypcap}

\usepackage[ruled,noline,noend,linesnumbered]{algorithm2e}

\newcommand{\au}{\,\mathrm{au}}

\newcommand{\gsim}{\lower.7ex\hbox{$\;\stackrel{\textstyle>}{\sim}\;$}}
\newcommand{\lsim}{\lower.7ex\hbox{$\;\stackrel{\textstyle<}{\sim}\;$}}

\newcommand{\D}{\emph{digest2}}
\newcommand{\DD}{D_2}
\newcommand{\MUK}{{\sc muk}}
\newcommand{\PG}{{\sc Pangloss}}
\newcommand{\pluseq}{\mathrel{+}=}

\newcommand{\NEOCP}{{\sc NEOCP}}

\shorttitle{The \D{} NEO Classification Code}
\shortauthors{Keys et al.}

\submitjournal{PASP, Day Month Year}

\begin{document}

\title{The \D{} NEO Classification Code}

\author{Sonia~Keys} 
\affiliation{Harvard-Smithsonian Center for Astrophysics, 60 Garden St., MS 51, Cambridge, MA 02138, USA}

\author[0000-0002-5396-946X]{Peter~Vere\v{s}} 
\affiliation{Harvard-Smithsonian Center for Astrophysics, 60 Garden St., MS 51, Cambridge, MA 02138, USA}
\correspondingauthor{Peter~Vere\v{s}}
\email{peter.veres@cfa.harvard.edu}

\author[0000-0001-5133-6303]{Matthew~J.~Payne}
\affiliation{Harvard-Smithsonian Center for Astrophysics, 60 Garden St., MS 51, Cambridge, MA 02138, USA}

\author[0000-0002-1139-4880]{Matthew~J.~Holman}
\affiliation{Harvard-Smithsonian Center for Astrophysics, 60 Garden St., MS 51, Cambridge, MA 02138, USA}

%
\author{Robert~Jedicke } 
\affiliation{Institute for Astronomy, 2680 Woodlawn Drive, Honolulu, HI 96822, USA}

\author{Gareth~V.~Williams}
\affiliation{Harvard-Smithsonian Center for Astrophysics, 60 Garden St., MS 51, Cambridge, MA 02138, USA}

\author{Tim~Spahr} 
\affiliation{NEO Sciences, LLC, 1308 Applebriar Lane, Marlborough, MA 01752, USA}

\author[0000-0003-0423-7112]{David~J.~Asher } 
\affiliation{Armagh Observatory and Planetarium, College Hill, Armagh, BT61 9DG, UK}

\author{Carl~Hergenrother } 
\affiliation{Lunar and Planetary Laboratory, University of Arizona, Tucson, AZ 85721, USA  }



\begin{abstract}
We describe the \D{} software package, a fast, short-arc orbit classifier for small Solar System bodies. 
The \D{} algorithm has been serving the community for more than 13 years. The code provides a score, $\DD{}$, which represents a pseudo-probability that a tracklet belongs to a given Solar System orbit type. 
\D{} is primarily used as a classifier for Near-Earth Object (NEO) candidates, to identify those to be prioritized  for follow-up observation. 
We describe the historical development of \D{} and demonstrate its use on real and synthetic data. 
We find that \D{} can accurately and precisely distinguish NEOs from non-NEOs. At the time of detection, 14\% of NEO tracklets and 98.5\% of non-NEOs tracklets have $\DD{}$ below the critical value of $\DD{}=65$.
94\% of our simulated NEOs achieved the maximum $\DD=100$ and 99.6\% of NEOs achieved $\DD{}\ge65$ at least once during the simulated 10-year timeframe.
We demonstrate that $\DD{}$ varies as a function of time, rate of motion, magnitude and sky-plane location, and show that NEOs tend to have lower $\DD{}$ at low Solar elongations close to the ecliptic.  
We use our findings to recommend future development directions for the \D{} code.
\end{abstract}


\section{Introduction}
\label{SECN:INTRO}

Near-Earth Objects (NEOs), defined as any small Solar System body having perihelion less than 1.3$\au$, are of significant interest for a number of reasons.  These include planetary defense, 
spacecraft missions, commercial development, and investigations into the origin and evolution of the Solar System.
The NEO discovery rate has risen rapidly in recent decades, driven by the 1998 Congressional mandate to discover 90\% of NEOs larger than 1\,km \citep[the Spaceguard Survey][]{Morrison92}, and the subsequent 2005 Congressional mandate to discover 90\% of NEOs larger than 140\,m (George E. Brown, Jr. NEO Survey Act \footnote{Section 321 of the NASA Authorization Act of 2005 (Public Law No. 109-155)}). 

Solar System objects are typically identified through their apparent motion in a series of images spanning minutes to hours. 
Given such a sequence of exposures, it is straightforward to identify ``tracklets''~\citep{Kubica07}, sets of detections that are consistent with an object with a fixed rate of motion. 
A related term is  an \textit{arc}, which may contain any number of tracklets, possibly from different observatories. 
The time span from first to last observation is termed the \textit{arc-length}.

It is straightforward to determine whether a tracklet represents a known object with a well defined orbit.  In that case, the observations coincide with the predicted positions for the known object.  Otherwise, the tracklet likely represents a newly discovered object.  It is then of interest to determine what type of orbit that object has, and in particular, if it is a new NEO.  Given limited telescopic resources, it is not possible to re-observe all objects to immediately determine their orbits. It is therefore essential to identify which objects are more likely to be NEOs and prioritize those for further observation.  

Although a tracklet does not uniquely determine an orbit, characteristic sky-plane motion can be used to infer a possible orbit type \citep[e.g.][]{1996AJ....111..970J}.  
The tool currently used for this is \D{}\footnote{There is no known history of a digest1; this is not version 2 of a program. Also nothing is known about the origin or meaning of the name. The program name is simply \D{}.}. 
The \D{} code uses only a single motion vector, derived from a tracklet or short arc, to identify all possible elliptical orbits consistent with that motion.   This set of possible orbits is then divided into disjoint orbital categories. 
Given a population model that represents the number of solar system objects of each type, we can estimate the likelihood that the tracklet represent a member of each category.  

With a suitable threshold, \D{} can serve as an NEO binary classifier.   Tracklets with high $\D{}$ scores, $\DD{}$, are posted on the Near-Earth Object Confirmation Page (NEOCP\footnote{\url{http://www.minorplanetcenter.net/iau/NEO/toconfirm_tabular.html}}).  These objects are prioritized by the NEO follow-up community for additional observations.

In this manuscript we describe the \D{} code.  The code as implemented has been used by the community for a number of years. Our primary goal is to describe and document the major elements of \D{} and the practical consequences of its design and implementation.  Although we highlight possible areas of improvement, these are left for future work.

\section{Methodology}
\label{s:METHOD}

The observations from a short-arc tracklet directly constrain the position, $(\alpha, \delta)$, and motion, $(\dot\alpha,\dot\delta)$, of the asteroid in the sky, but the topocentric radial-distance, $\rho$, and radial-velocity, $\dot\rho$, between the asteroid and the observer are essentially unconstrained.   This is the fundamental challenge of orbit determination.

An \emph{assumed} $\rho$ allows a heliocentric position to be derived (see Appendix~\ref{app:admiss}), and with the assumption of a topocentric $\dot\rho$, a heliocentric velocity can also be calculated. 
\citet{Milani04} refer to the ``admissible region'' as being the set of $(\rho,\dot\rho)$ for which the resultant orbit is heliocentrically bound (elliptical w.r.t. the Sun). 

Many authors have addressed the problem of short-arc orbit determination and constraint \citep[e.g.][and discussions therein]{Vaisala39,Marsden85,Bowell89,1990acm..proc...19B,Marsden91,Tholen2000, Milani04,Milani05,Oszkiewicz09,Spoto18}.
Of particular relevance to \D{}, in the late 1980s and early 1990s, R. H. McNaught of the Siding Spring Observatory undertook an extensive effort to determine allowable orbital element ranges and class-specific object probabilities based on two observations and a magnitude~\citep[\PG{}][]{McNaught99}. 
The method proceeded by stepping through a range of possible topocentric distances and angles (between the velocity vector and the line of sight) which were consistent with the observations and bound to the sun. 
\PG{} used a population model to assign weights to different orbit classes.  
The \PG{} code provided the foundation for the \D{} code described in this manuscript.   

We review two related methods for addressing the problem of short-arc orbits.
The first concerns the technique referred to as ``statistical ranging'' \citep{Virtanen01}, in which two observations are selected from the tracklet, thus fixing $(\alpha,\dot\alpha,\delta,\dot\delta)$, and then topocentric ranges at each of the epochs of the two observations are randomly chosen and a corresponding orbit computed from the admissible region that is compatible with the observational tracklet data. 
This process is repeated over many random topocentric distances to generate a set of compatible orbits.

The second method refers to ``systematic ranging''.  This method was introduced by \citet{Chesley05} and is described in detail in \citet{Farnocchia15}.  In this method, a systematic raster over $(\rho,\dot\rho)$ space is performed, generating an orbit for each and comparing to \emph{all} tracklet observations. 
The RMS of the fit residuals for each $(\rho,\dot\rho)$ point indicates the quality of the fit, and $\chi^2$ probabilities can be used to derive confidence regions in $(\rho,\dot\rho)$ space.

While \D{} is not directly derived from either \citet{Virtanen01} or \citet{Chesley05} (and the \PG{} ranging code, from which \D{} is derived, predates both), the \D{} code uses the same fundamental approach that is common to both statistical and systematic ranging techniques: sets of bound heliocentric orbits are generated that all satisfy the short-arc observations.

\subsection{Historical Development of \D{}} 
\label{s:HIST}
As described above, the early origins of the \D{} were the \PG{} code developed by R. McNaught.
The current version of \D{} evolved from a FORTRAN 77 code (``\texttt{223.f}'') employed in \citet{1996AJ....111..970J}, and then further developed by C. Hergenrother and T. Spahr from the \PG{} code of R. McNaught.  
It accepted observation files in the MPC's 80-character format\footnote{https://minorplanetcenter.net/iau/info/ObsFormat.html},  encoded the vector solution described in section~\ref{s:SolnAlgo}, employed the parabolic limit described in appendix \ref{app:admiss}, and used the model population look-up described in section~\ref{s:POPN}.  

Since the first publication,\D{} has undergone many incremental improvements.  In Appendix \ref{app:HIST}, we provide a detailed description of the key changes to the \D{}.


\section{Computational elements of \D{}}
\label{s:COMP}
In this section we describe the main components of the current version of \D{} and explain the algorithmic choices in its development. 
We emphasize that we are providing a factual report of how the code was implemented, and hence how the code has been operating for more than a decade. 
Significant improvements are possible in future versions, but the focus of this paper is documenting the development, features, and performance of the existing code. 

We discuss \D{}'s key algorithmic steps in Sections \ref{s:EndPoint} to \ref{s:score}, and then summarize them in Section \ref{s:AlgSumm} and Algorithm Table~\ref{alg: essential}.

\subsection{Endpoint Synthesis}
\label{s:EndPoint}

\begin{figure}[ht]
\centering
    %
    %
  \includegraphics[trim = 0mm 0mm 0mm 0mm, clip, angle=0, width=0.49\textwidth]{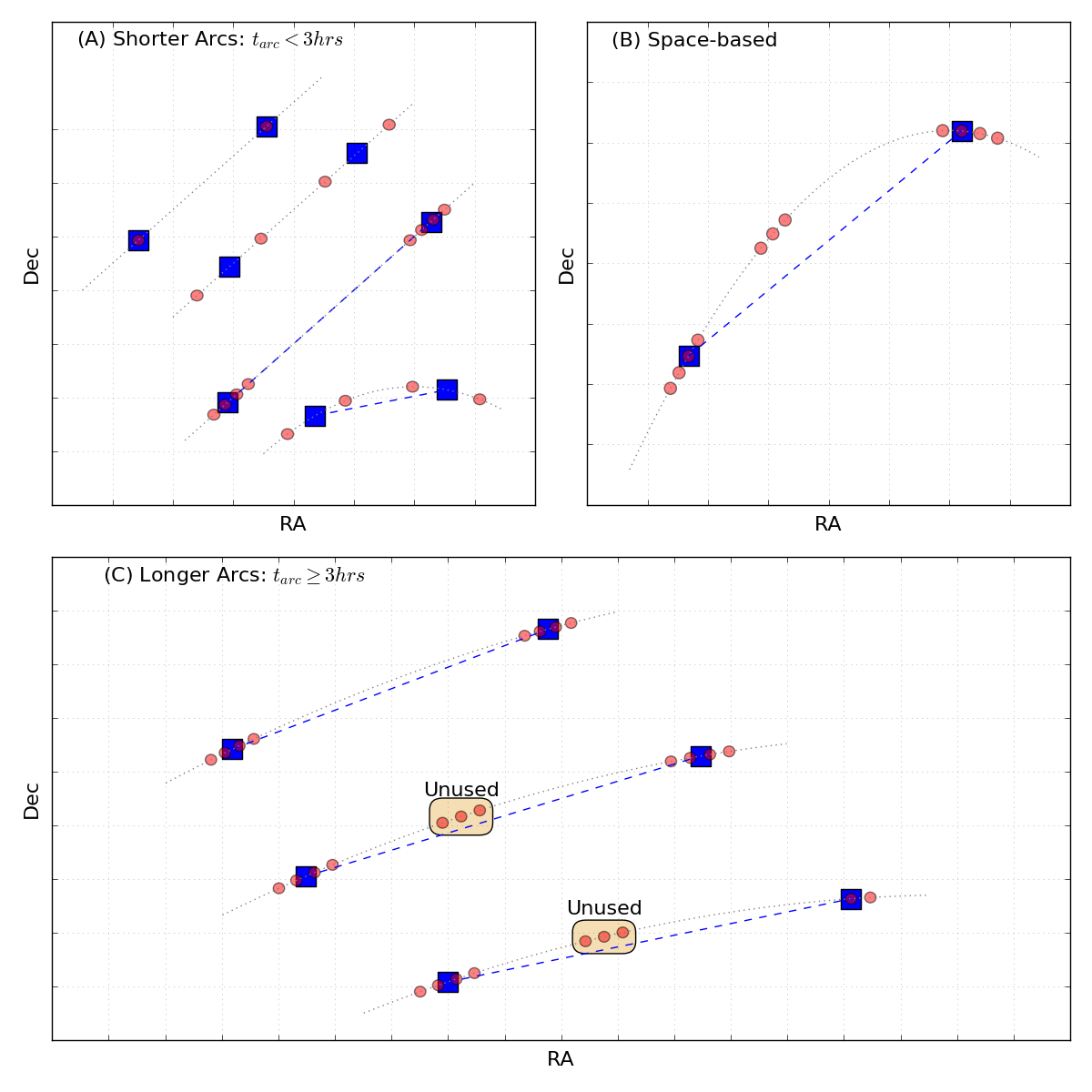}
  \caption{%
    Schematic illustration of the construction of two-element tracklets within \D{} for short-arcs ($<3\,hrs$: top-left), space-based data (top-right), and long-arcs ($\geq3\,hrs$: bottom). 
    Detections are plotted as red circles along their underlying path (gray dotted line);
    The two-element tracklets constructed by \D{} are the blue squares and dashed lines.
    A detailed description of the algorithm used to construct the two-element tracklets is provided in the text of \S\ref{s:EndPoint}.
    }
    \label{fig:tracklet}
\end{figure}

The algorithm starts by reading tracklet and population-model data from input files.
Since version 11 (June 2011), \D{} considers all observations in the arc and synthesizes end points to define a motion vector, with the aim of improving the robustness of \D{} against both bad observations and statistical observational error. 
Figure~\ref{fig:tracklet} demonstrates a number of cases handled by the code.

{\it Short-Arcs: }
Panel A illustrates relatively short arcs, with all observations from the same site.  
At the top, both observations in the two-observation-arc are used as-is. 
For the second arc, a more typical tracklet with four observations, a great circle linear motion vector is fit to the observations.
Endpoints are synthesized near the 17th and 83rd percentile of the observations in the arc, instead of near the extreme endpoints. This is not to account for positional uncertainty, but rather is an approximate method to account for cases in which excessive measurement error or poor astrometry can significantly shift the endpoint, and thus change the shape of the admissible region.\footnote{The justification for the particular choice of 17th and 83rd percentiles is unknown to the surviving authors. Moreover, we emphasize that future versions of the \D{} code will improve on the handling of measurement uncertainty and tracklet construction.}.  
For the third arc, synthesizing endpoints near the 17th and 83rd percentile yields synthetic observations within the tracklet.  
The fourth arc shows significant curvature, with the synthesized linear motion vector exhibiting significant residuals w.r.t all observations.

{\it Space-based observations\footnote{{\bf \url{https://minorplanetcenter.net/iau/info/SatelliteObs.html}}}: }
As illustrated in Panel B, space-based observations often show strong curvature on the sky due to parallax as the spacecraft orbits the earth.  
In this case interpolation on a great circle fit is not meaningful.  
If any observations in an arc are space-based, two actual observations are selected, near the 17th and 83rd percentile.

{\it Longer-Arcs: }
Panel C shows cases of longer arcs and/or arcs with multiple observing sites.  
Here, two sub-arcs are selected where each sub-arc is < 3 hours and all detections in a sub-arc are from the same site.  These two sub-arcs are then separately reduced to yield the motion vector end points.  In the top arc of Panel C, sub-arcs can be selected to use all available observations.  A separate great circle linear motion fit is then applied to each sub-arc and observations are synthesized, again near the 17th and 83rd percentiles.  
The second arc illustrates how the conditions that sub-arcs be < 3 hours and all from the same site may leave a number of observations unused.  
The third arc shows a case where the 83rd percentile is outside of the end sub-arc.  
In this case no point is synthesized outside the sub-arc, rather the first observation of the sub-arc is used as-is.

\subsubsection{Dithering and Observational Uncertainty}
\label{s:Dithering}
%
\begin{figure}[htp]
\centering
    %
    %
  \includegraphics[trim = 0mm 0mm 0mm 0mm, clip, angle=0, width=0.49\textwidth]{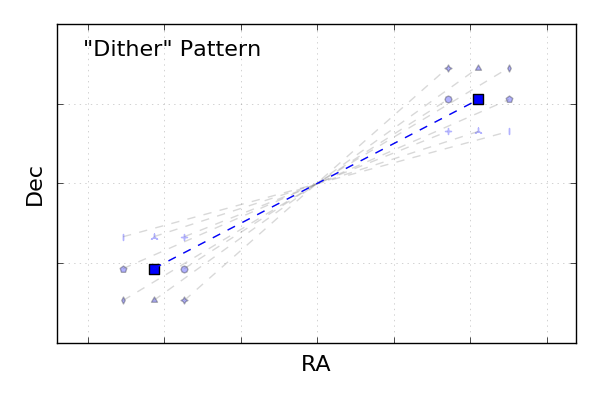}
  \caption{%
    Schematic illustration of the construction of ``dithered'' end-points to a two-point tracklet constructed as per Figure \ref{fig:tracklet}, Section \ref{s:Dithering}.
    The nominal end-points (large blue squares) are varied within the uncertainties, $\sigma$, of the input detections by adding $\pm0.5\sigma$ to one-or-both of the RA and Dec values of the end points, producing 8 additional \emph{deterministic} variations (small gray symbols).
    We refer the reader to the main text of Section \ref{s:Dithering} for a brief discussion of the short-comings of this approach. 
    }
    \label{fig:dither}
\end{figure}
%
Once a two-point tracklet has been constructed as described in Section \ref{s:EndPoint}, the end-points are varied within the uncertainties, $\sigma$, of the input detections.
The nominal end-points are varied by adding $\pm0.5\sigma$ to either (or both) of the RA and Dec values of the initial end points, producing a total of 9 variant tracklets: the initial one, plus the additional 8 variations illustrated in Figure \ref{fig:dither}.

The astrometric uncertainties used for $\sigma$ can be defined per observatory code in the \D{} configuration file. 
The current MPC settings of uncertainties and keywords are presented in Table~\ref{t:tabUnc} of Appendix \ref{app:tables}. 

While this approach has the effect of producing a range of different motion vectors, it should be emphasized that this method of dealing with astrometric uncertainty is \emph{not} ideal from a statistical viewpoint, omitting occasional astrometry that is worse than our statistics, or significantly better due to improved astrometric catalog or measurement. 
Improvements expected in the near future will employ astrometric uncertainties directly reported with the submitted astrometry on an individual detection basis to handle observational uncertainty.

\subsection{Orbit Solution}
\label{s:SolnAlgo}

Given a nominated observer-object distance, $D$, and a nominated angle, $\alpha$, between the observer-object unit vector, $\vec{d}$, and the object velocity vector, a specific orbit can be computed (Appendix \ref{app:admiss}).  
The $H$-magnitude can be calculated for each orbit using the geocentric vector $\vec{\Delta}=D\,\vec{d}$ and the heliocentric distance of the object, $\vec{r}$, to get the object phase angle $\Phi$, which, with the apparent magnitude $V$, gives $H$ in the IAU adopted H-G magnitude system \citet{1989aste.conf..524B}.

\subsection{Population Model used by \D{}}
\label{s:POPN}
The \D{} code requires a population model against which tracklet-derived-orbits can be compared, representing the number of solar system objects in a variety of dynamical categories.  
\D{} uses two different population models.  
The first is a full population model including all objects down to a given diameter. 
The second is the difference between the full population model and the known orbit catalog, representing the undiscovered population.

\begin{figure*}[ht]
\begin{minipage}[b]{\textwidth}
\begin{verbatim}
Desig.    RMS Int NEO N22 N18 Other Possibilities
 NE00030  0.15 100 100  36   0
 NE00199  0.56  98  98  17   0 (MC 2) (JFC 1)
 NE00269  0.42  24  23   4   0 (MC 7) (Hun 3) (Pho 15) (MB1 <1) (Han <1) (MB2 41) (MB3 5) (JFC 1)
\end{verbatim}
\caption{Sample output from the \D{} code showing scores for 3 tracklets. Columns from left: Designation of a tracklet, root-mean-square (RMS) of great-circle-fit of the tracklet in arcseconds, and $\DD{}$ for the orbit classes (Int, NEO, N22, N18, Other Possibilities) described in Table~\ref{t:classes} of Appendix~\ref{app:tables}.}
\label{f:verbatimA}
\end{minipage}
\end{figure*}

\subsubsection{Full Population Model}
\label{s:POPN:Full}
\D{} uses the term ``raw'' to indicate when the full population model is used to score tracklet-derived orbits. 
The current version of \D{} uses the Pan-STARRS Synthetic Solar System Model (S3M) of \citet{Grav11}, consisting of over 14 million simulated Keplerian orbits.  

\D{} bins S3M into 15 different orbit classes (Table \ref{t:classes} in Appendix~\ref{app:tables}). 
The model uses bins in perihelion ($q$), eccentricity ($e$), inclination ($i$) and absolute magnitude ($H$).  
Binning in perihelion, rather than semi-major axis $a$, enables a bin cut at $q = 1.3$ to directly distinguish NEO orbits from non-NEO orbits.
The binning is non-uniform and provides enhanced resolution in higher density regions of parameter space, while simultaneously reducing the number of empty and near-empty bins in low density regions.
The bin boundaries and counts for the full population are depicted in Figures~\ref{fig:Population} in Appendix~\ref{app:bins}. The binned population and model reduction is done by the  \MUK{} program\footnote{\MUK{} is available from \url{https://bitbucket.org/mpcdev/d2model/src/master/muk.c}}.

\subsubsection{Unknown Population}
\label{s:POPN:Unknown}
\D{} also allows a comparison of tracklet-derived orbits against the likely population of ``undiscovered'' or ``unknown'' objects, referred to as the ``no ID'' model. 
An argument can be made that ``unknown'' objects can be more accurately scored against a population model that excludes objects typical of those that are already known.  

Given a binned version of a full population model (as described in Section \ref{s:POPN:Full}), the desired ``unknown'' population model is constructed by reducing bin populations by the number of cataloged objects with well determined orbits.  
Given a metric for orbit quality, an orbit is selected from a catalog \emph{of known objects} and the orbit is graded as to whether it would likely be identified. If the catalog orbit is identifiable, the count of objects in the appropriate bin within the \emph{full, modelled} population can then be reduced.
Repeating this procedure for all well-known objects reduces the full population model to the ``unknown'' population model. This model reduction and bin selection is performed by the above-mentioned program \MUK{}.

The known catalog currently used by \MUK{} is \texttt{astorb.dat}\footnote{\texttt{astorb.dat}, and is available from \url{ftp://ftp.lowell.edu/pub/elgb/astorb.dat.gz}} \citet{1994IAUS..160..477B}, \citet{astorb}. Its orbit quality metric is \texttt{astorb} field 24 \citep{1993Icar..104..255M}, corresponding to the peak ephemeris uncertainty over a ten year period. 
The Peak ephemeris uncertainty is observationally motivated, offers a direct comparison regarding how secure the object is against getting lost, and is the most generic way to compress the information of the covariance matrix into a
scalar. 
\MUK{} considers such an orbit to be determined well enough that its identification with a tracklet would be trivial.  For each of these orbits, the corresponding model bin is decremented by one, unless the bin population reaches zero.

A binned population model divided into bins for ``raw'' and ``no id'' populations is distributed with the \D{} source code and updated when a new version of \D{} code is available.

\subsection{Searching for bins}
\label{s:bin search}

Given an orbit calculated as described in Section \ref{s:SolnAlgo}, the next step is to assign the orbit to the appropriate $(q,e,i,H)$ population-bin. The algorithm searches over a range of distances, $D$, and angles, $\alpha$, within the admissible region (Section \ref{s:METHOD}). 
There are a number of possible approaches to selecting a set of points for evaluation within this region. 
\D{} does \emph{not} use a $\chi^2$ (or RMS) approach, nor does it generate a fixed grid of points, but instead utilizes a binary search approach (See Appendix~\ref{app:bin_search}).

\subsection{Class Score}
\label{s:score}

Given the set of bins identified in Section \ref{s:bin search}, a ``class score'' is calculated over those bins. 
These scores are calculated for any of the classes listed in Section \ref{s:POPN} that are selected by the user at run-time as being of interest.

For a specified orbit class of interest, a sum $\Sigma_{class}$ is accumulated based on the modeled population consistent with the orbit class.  
Another sum, $\Sigma_{other}$, is accumulated for the modeled population consistent with orbits {\it not} of the class.  
The class score $S_c$ is given by
\begin{equation}
\label{score_function}
S_c = 100 \frac{\Sigma_{class}}{\Sigma_{class} + \Sigma_{other}}
\end{equation}

Therefore, the \D{} score, $\DD{}$, ranges from 0 to 100 in any given class. The example in Figure \ref{f:verbatimA} demonstrates a range of $\DD{}$ values for three tracklets of different objects, after execution of the compiled $\D{}$ program\footnote{\url{https://bitbucket.org/mpcdev/digest2/overview}} with its supplied sample input files\footnote{\url{https://bitbucket.org/mpcdev/digest2/downloads/}}.

The output generated by \D{} lists scores for tracklets on individual rows. The number of displayed columns as well as the configuration of \D{} can be controlled using keywords in the configuration file (See Table \ref{t:keywords} of  Appendix~\ref{app:tables}). 
The first object (NE00030) appears to be an NEO and an ``Interesting'' object based on $\DD{}=100$. However, its scores for being a large NEO ($N22$, $N18$) are rather low, so this object is therefore most likely a small NEO with $H>22$. 
The second object ($NE00199$) also has large NEO scores, however, its RMS is large, suggesting either astrometry with large errors or deviation from a great circle motion due to a close encounter with the Earth. 
The last two columns, in the parentheses, suggest that there is a small probability that this object is a Mars Crosser or a Jupiter Family Comet. 
The third object ($NE00269$) also has a large RMS and its NEO scores are low ($\DD{}<25$). Based on other orbit classes computed for $NE00269$, this object is most likely a Central Main-Belt asteroid.

\subsection{Algorithmic Summary}
\label{s:AlgSumm}

In Algorithm Table \ref{alg: essential}, we summarize the key algorithmic steps employed in \D{}, These are:
(1) the reduction of the tracklet observations to a single motion vector (Line \ref{ess:two pos}), 
(2) the sampling of the distance, $D$, and velocity angle, $\alpha$ (Lines \ref{ess:dist loop}-\ref{ess:angle loop}), 
(3) the generation of orbital elements and population comparison (Lines \ref{ess:elements}-\ref{ess:pop}), 
leading to 
(4) the accumulation of a score (Lines \ref{ess:accumulate}-\ref{ess:score}).

\begin{algorithm}
\caption{\D{} essential algorithm\\See Appendix~\ref{app:admiss} and~\ref{app:bin_search} for more details. 
}
\label{alg: essential}
\DontPrintSemicolon
\SetKwFor{ForEach}{for each}{do}{}
    Open observation file\;  \label{ess:load_obs}
    Load binned population model into memory\; \label{ess:load_popn}
    \ForEach{\upshape $arc$ of identical $designations$ in file}{ \label{ess:read arc}
        Select or derive two positions from $arc$\; \label{ess:two pos}
        Compute a magnitude $V$ from $arc$\; \label{ess:V mag}
        Initialize score accumulators\;
        \For{\upshape a number of nominated distances $D$}{ \label{ess:dist loop}
            Compute distance dependent vectors $\{\vec{v_D}\}$\; \label{ess:dist vec}
            Compute angle limits $\alpha_1, \alpha_2$ from  $\{\vec{v_D}\}$\; \label{ess:angle lim}
            Compute $H$ from $V$ and $\{\vec{v_D}\}$\; \label{ess:H mag}
            \For{\upshape nominated angles $\alpha$ in range $\alpha_1, \alpha_2$}{ \label{ess:angle loop} 
                Compute state vector at $\alpha$\; \label{ess:state vec}
                Compute orbital elements from state vector\; \label{ess:elements}
                Look-up bin population for elements\; \label{ess:pop}
                Accumulate values needed for scoring\; \label{ess:accumulate}
            }
        }
        Compute $score$ from accumulated values\; \label{ess:score}
        Output arc results, minimally $designation$ and $score$\;
    }

\end{algorithm}

As described in Section \ref{s:SolnAlgo}, \D{} generates a large number of different observer-object distances ($D$) and angles ($\alpha$)  between the observer-object unit vector and the object velocity vector, each pair of $(D,\alpha$) yielding different keplerian orbital elements.
The distance, $D$, and angle $\alpha$, are selected in order to construct a bound heliocentric orbit.

For each candidate orbit, the elements index a bin (Section~\ref{s:bin search}) of a population model of the Solar System (Section~\ref{s:POPN}). 
Indexed bin populations then represent populations of modeled objects consistent with the input arc.  
For a dynamic class of interest, such as NEOs, population sums can be accumulated  that allow score calculation (Section~\ref{s:score}).  
A score indicates a quasi-likelihood that the input arc represents an object of the class of interest.

\section{\D{} score analysis}
\label{s:Classifier}

The primary use of the \D{} code has been as a \emph{binary classifier} for NEOs, in which a tracklet's NEO $\DD$ score\footnote{Unless otherwise indicated, in the remainder of this work we will understand $\DD{}$ to refer to the NEO "no id" (NID) \D{} score.} is compared to a critical value, $D_{2,crit}$\footnote{Since 2012, $D_{2,crit}=65$.} 
Tracklets with $D_{2} > D_{2,crit}$ are considered by the MPC to be eligible for submission to the Near-Earth Object Confirmation Page (NEOCP). 
The rationale and history for selecting a given score is described later at the end of Section~\ref{SECN:RES:ACCURACY}.
Submitting newly discovered tracklets to the NEOCP allows them to be prioritized for follow-up by the community, allowing their arcs to be extended, and a more precise orbit to be determined for them, ultimately determining whether the object is indeed an NEO.

The value of $D_{2,crit}$ used to decide whether a tracklet is admissible to the NEOCP has varied since the introduction of \D{}.
This has changed both the \emph{number} of tracklets admitted to the NEOCP and the \emph{purity} of the tracklets (what fraction are NEOs) on the NEOCP.
The detailed effects of these changes are discussed at length in Section \ref{SECN:RES:ACCURACY}.

\subsection{Application of \D{} to Individual Objects}
\label{s:IND}
%

\begin{table}
\small
\begin{center}
\footnotesize
\caption{Individual NEOs (top) and non-NEOs (bottom) selected for detailed study in Section \ref{s:IND}.}
\tabcolsep=0.11cm
\begin{tabular}{l|c|c|ll|ccc}
\hline
                         & &  &  &  & $q$&$e$&$i$ \\
Designation & H & PHA & \multicolumn{2}{c|}{Orbit type} & $[au]$&$-$&$[deg]$ \\
 \hline
 \hline
(198752) & 19.8  & No &NEO&Amor         & $1.21$&$0.52$&$1.75$\\
2015 DP155 & 21.5& Yes &NEO&Amor        & $1.02$&$0.22$&$5.38$\\
2012 HN40 & 20.4 &  No &NEO&Apollo      & $0.89$&$0.67$&$14.39$\\
2012 HZ33 &  20.4 &Yes &NEO&Apollo     & $0.95$&$0.21$&$23.88$\\
\hline
\hline
2006 EW1 & 16.0 & No    &non-NEO&MB    & $2.80$&$0.12$&$6.03$ \\
2000 ED68 & 16.4  & No  &non-NEO&MB     & $1.89$&$0.40$&$25.40$   \\
2013 BX45 & 17.6 & No   &non-NEO&MC     & $1.56$&$0.40$&$29.51$ \\
2017 BG123 & 19.5 &No   &non-NEO&HUN    & $1.60$&$0.13$&$23.29$ \\
2014 QC158 & 15.9& No   &non-NEO&HIL    & $3.12$&$0.22$&$14.19$ \\
2012 RW6 & 13.1 &  No   &non-NEO&TRO    & $4.87$&$0.06$&$16.24$ \\
\hline
\hline
\end{tabular}
\label{Digest_nonNEOs}
\end{center}
\end{table}
%
We provide illustrative examples of the \D{} score for a number of representative objects from the MPC database.
The objects and their orbital classifications are listed in Table \ref{Digest_nonNEOs}.

\begin{figure}[htb]
\centering
  \includegraphics[trim = 0mm 30mm 10mm 40mm, clip, angle=0, width=0.8\columnwidth]{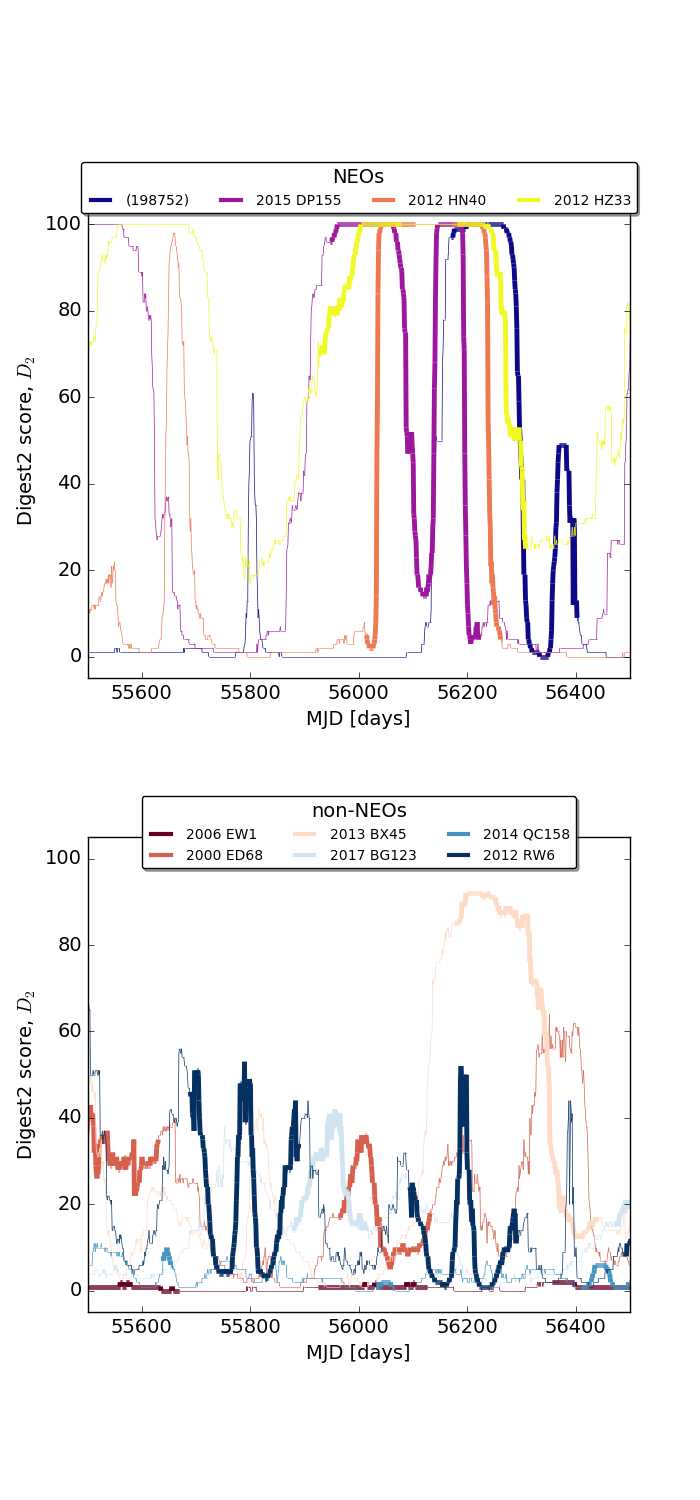}
  \caption{%
    {\bf Top:} Examples of the variation in \D{} score as a function of time for the NEOs in Table \ref{Digest_nonNEOs}.
    {\bf Bottom:} Examples of the variation in \D{} score as a function of time for the non-NEO objects in Table \ref{Digest_nonNEOs}.
    Thick lines indicate when the objects were practicably observable. 
    All plotted NEOs have a maximum \D{} score $\sim 100$, but all exhibit periods during which their scores are significantly lower, and all spend some time during which the score is $<65$.
    The sample non-NEOs display a wide range of behaviors. 
    Some, such as the MBA ``2006 EW1'' always display low \D{} scores. 
    In contrast, some objects such as the Mars-crosser ``2013 BX45'' spend sizable fractions of the time with a score $>65$.
    In the remainder of Section \ref{s:Classifier} we examine the primary drivers of this variation. 
    \label{fig:individual}
    }
    
\end{figure}

\begin{figure*}[htb]
\begin{minipage}[bht]{\textwidth}
\centering
  \includegraphics[trim = 0mm 0mm 0mm 0mm, clip, angle=0, width=\textwidth]{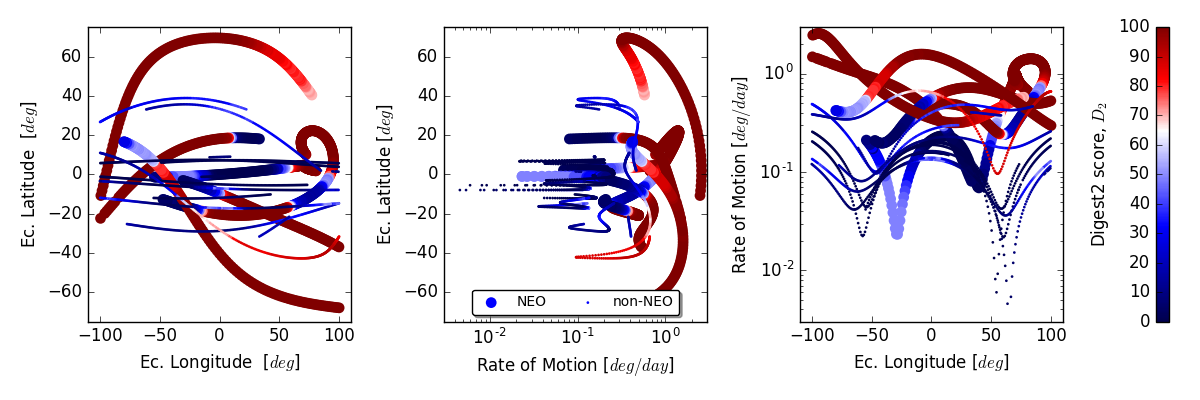}
  \caption{%
    Colormaps of the variation in \D{} score as a function of opposition-centered ecliptic latitude, longitude and the rate of motion during the times of visibility of the individual objects from Figure \ref{fig:individual}.
    The large diameter circles are the NEOs, and the small points are the non-NEOs. 
    It is clear from the center plot that the value of the NEO \D{} score is strongly dependent on the rate-of-motion and latitude. %
    }
    \label{fig:IndHeat}
\end{minipage}
\end{figure*}
%
Because the observing cadence for object tracklets submitted to the MPC is rather sparse, we generated daily ephemerides and synthesized tracklets for our sample over a period of 10 years. 
For each object, the orbit was numerically integrated and an ephemeris was generated for two epochs per night, separated by 20 minutes.
The pair of points within a night is then transformed into Geocentric (RA,Dec) coordinates, and used to define a tracklet for each night. 
Each tracklet is then used as an input to the \D{} code, and a score generated. 

All of the NEOs plotted in Figure \ref{fig:individual} have a maximum \D{} score $\sim 100$, but all exhibit periods during which their scores are significantly lower, and all spend some time during which their score is $<65$.
The \emph{non}-NEOs in Figure \ref{fig:individual} have a wide range of behaviors. 
Some, such as the MBA ``2006 EW1'' always display low \D{} scores. 
In contrast, some objects such as the Mars-crosser ``2013 BX45'' spend sizable fractions of the time with a score $>65$.

Because a critical \D{} score $D_{2,crit}=65$ is  used by the MPC to post objects to the \NEOCP{} and prioritize them for observational follow-up, it follows from Figure \ref{fig:individual} that, 
(a) an NEO could fall below $D_{2,crit}$ for some fraction of the time it is visible, and hence ``miss-out'' on being sent to the \NEOCP{},  
and 
(b) non-NEOs can gain scores higher than $D_{2,crit}$, and hence be ``unnecessarily'' sent to the \NEOCP{}. 
In the following sections we quantify the frequency of such scenarios. 

In Figure \ref{fig:IndHeat} we plot $\DD{}$ as a function of ecliptic coordinates (with respect to opposition) and the rate of motion. The plotted positions are for the time of visibility depicted in Figure \ref{fig:individual}.
It is clear from the middle plot of Figure \ref{fig:IndHeat} that the value of the \D{} score is strongly dependent on the rate-of-motion and latitude. 
While this \emph{generally} serves to distinguish NEOs from non-NEOs, we can see that some of the non-NEOs spend some fraction of their time with relatively high rates-of-motion and/or high-ecliptic latitudes. 
This inevitably leads to confusion between NEO and non-NEO, as evidenced by the high $\DD{}$ scores for some of the non-NEOs.

\subsection{Application of \D{} to Populations}
\label{s:POP}

\subsubsection{Data Sets and Simulations}\label{s:Sets}
\begin{deluxetable}{lllll}
  \caption{Description of population data sets used in this work. The $MPC_{D,11}, MPC_{D,17}$ and $MPC_{A,17}$ data sets contain real observations, the $Sim$ set contains synthetic observations based on real orbits, while $LSST$ and $Granvik$ contain synthetic observations based on synthetic orbits.}
  \label{t:Data_Sets}
  \tablehead{\colhead{Data Set} & \colhead{Type} & \colhead{Timespan}  & \colhead{Tracklets} & \colhead{Objects}}
  \startdata
  $MPC_{D,11}$ & Real (All) & 1-Year  & 14,003&14,003\\
  $MPC_{D,17}$ & Real (All) & 1-Year  &18,715& 18,715\\
  $MPC_{A,17}$ & Real (All) & 1-Month   &374,406&143,983\\
  $Sim$ & Synthetic (NEO) & 10-Years &$3.8\times10^6$&16,230 \\
  LSST & Synthetic (NEO, MB) & 9-days &$14.3\times10^6$&$1.4\times10^6$\\
  Granvik & Synthetic (NEO) & 1-Year &618,951&14,458 \\
  \hline
  \enddata
\end{deluxetable}

To investigate the behaviour of $\DD{}$ for a large population, we computed $\DD{}$ for the data sets listed in Table~\ref{t:Data_Sets}. 
The data sets $MPC_{D,11}$ and $MPC_{D,17}$ contain all \emph{discovery} tracklets\footnote{A \emph{discovery} tracklet is the first reported tracklet of an object} submitted to the MPC in 2011 and 2017 respectively, thus containing tracklets of all types of orbit. 
We use both $MPC_{D,11}$ and $MPC_{D,17}$ data sets because in 2011 the MPC used $D_{2,Crit}=50$ for posting to the NEOCP, whereas in 2017 the MPC used $D_{2,Crit}=65$, allowing us insight into the effects of different value of $D_{2,Crit}$ used in the past.

$MPC_{A,17}$ contains a month's worth of \emph{all} tracklets submitted to the MPC, containing all types of submitted orbit.
We selected January 2017 as a time when Jupiter's Trojans were observable near opposition.
We only selected tracklets fainter than 18 magnitude, to work in the regime of newly discovered objects.

The $Sim$ data set was obtained by integrating the known catalog of NEOs (as of July 2018) with $H>18$ for 10 years, starting on January 1, 2008 and deriving a daily ephemeris from the Geocenter, providing a two-detection tracklet spanning 1-hour for $\DD{}$ computation. To simulate accessibility for ground-based telescopes, we constrained the maximum V-band magnitude to 22.5 and distance from opposition to be within $\pm100$ degrees in longitude and $\pm70$ degrees in latitude. 90\% of objects in the catalog were observable within the $Sim$ data. 

The $LSST$ data set contains tracklets from 9-nights worth of a simulated LSST survey \citep{Veres17}. The data set contains NEOs from \citet{Granvik18} and Main-Belt asteroids from \citet{Grav11}.

The $Granvik$ data set contains NEO tracklets for the $H>18$ objects in the synthetic population of \citet{Granvik18} that we propagated for one calendar year and then used the resulting ephemerides to compute the value of $\DD{}$ on a nightly basis. 
We employed the same observability criteria for the limiting V-band magnitude and the opposition-centered ecliptic coordinates as we used on the $Sim$ data set.

\subsubsection{Distribution of \D{} Scores}
\label{s:DIST}
%
\begin{figure}[htp]
\centering
  \includegraphics[ width=0.9\columnwidth]{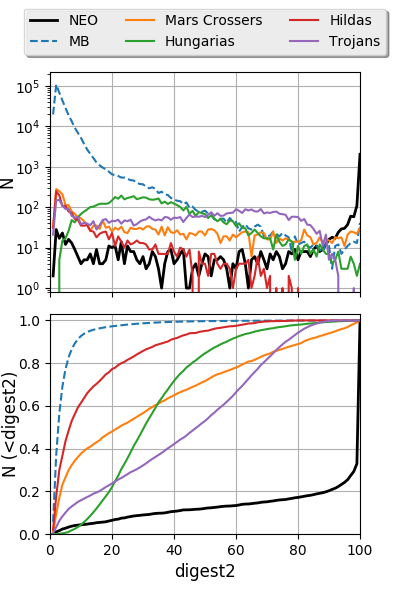}
  \caption{
  {\bf Top:} Distribution of \D{} scores for 1-Month's worth of all tracklets submitted to the MPC ($MPC_{A,17}$). 
  {\bf Bottom:} Cumulative version of the top plot. The majority of NEOs have $\DD{}\sim100$, while Main-Belt objects tend to have $\DD{}\approx0$. 
  Trojans are particularly numerous for $\DD{}\subset(55-90)$.}
\label{fig:month_A}
\end{figure}

The overall distribution of scores for the $MPC_{A,17}$ data set can be seen in Figure~\ref{fig:month_A}. 
This data set contains various object types. 
We plot histograms of the $\DD{}$ score at the top, and fractional cumulative distributions below. 
Most Main-Belt asteroids have $\DD\sim0$, and most NEOs have $\DD\sim100$. However, the long tail of NEOs with low $\DD{}$ is interesting because those with $\DD<D_{2,Crit}$ are not recognized as NEOs and will not be placed onto the NEOCP for follow-up. 
In the $MPC_{A,17}$ data set, 14\% (98.5\%) of NEO (non-NEO) tracklets had $\DD<D_{2,Crit}$.
Figure~\ref{fig:month_A} shows that Trojans and Mars-Crossers are objects that ``mimic NEOs'', as these two populations of object have the highest fraction with $\DD>D_{2,Crit}$. 
Trojans dominate numerically for $55<\DD<90$,
while Mars-Crossers contribute equally across the entire range of $\DD{}$ values.

\subsubsection{Comparison of NEO \D{} Scores Across Data Sets}
\label{s:NEOs}

\begin{figure}[htp]
    \label{f:NEO_d2}
    \includegraphics[width=\columnwidth]{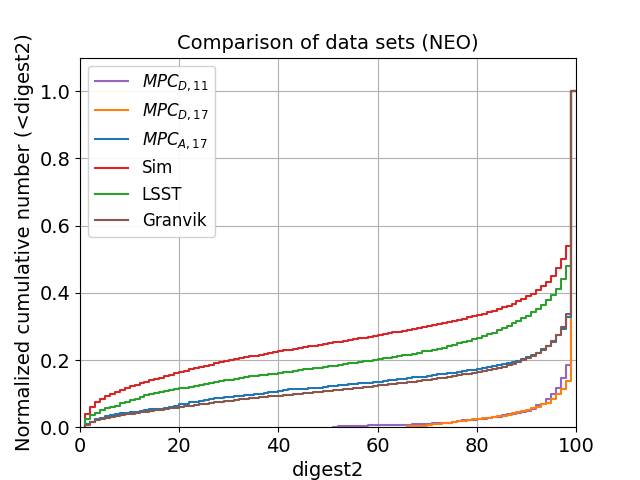}
    \caption{ Cumulative distribution of \D{} scores for all \emph{NEO} data sets. 
    The blue ($MPC_{A,17}$) line is reproduced from Figure \ref{fig:month_A}.
    While most NEO tracklets have $\DD{}=100$ at any given time, the distribution of smaller $\DD{}$ differs between data sets. 
    In the unbiased $Sim$ sample about 30\% of NEO tracklets have $\DD{}<D_{2,Crit}$.
    }
\end{figure}
%
Figure \ref{f:NEO_d2} shows the cumulative distribution of $\DD{}$ for the NEOs generated within the data sets of Table~\ref{t:Data_Sets}. 
The data sets differ at small $\DD{}$, introducing differing biases. 
The \emph{discovery} data sets ($MPC_{D,11}$, $MPC_{D,17}$) have the smallest fraction of tracklets in the low $\DD{}$ range and clearly display the effects of imposing a $D_{2,Crit}$-cut-off on objects submitted to the NEOCP.

The data sets containing \emph{all} submitted tracklets ($MPC_{A,11}$) or simulated tracklets ($Granvik$, $LSST$, and $Sim$) have increased ratios of low-scoring NEOs because (a) they are not subject to the $D_{2,Crit}$ threshold placed on objects sent to the NEOCP, and (b) they contain objects observed repeatedly.
The largest fraction of low scoring tracklets comes from the $Sim$ data set, which may be a consequence of the long-term simulation window (10-years) and our optimistic nightly cadence over the entire visible night sky, allowing it to contain objects with large synodic periods, and objects that only occasionally have $\DD>D_{2,Crit}$. 
The $Sim$ data set has a wide range of NEOs: 28\% of all visible tracklets had $\DD{}<65$ at any given time.
However, we note that within the data set, some NEOs are visible  for only a single one night, while others are visible for many months.

\subsubsection{Detailed Analysis of \D{} scores for NEOs}
\label{s:NEOs_details}
We use the $Sim$ data set to examine the frequency with which NEOs and their subclasses 
(Amor, Apollo and Aten type orbits, PHAs\footnote{Potentially Hazardous Asteroids have MOID $\leq0.05\au$ and $H\geq22$.} and low-MOID\footnote{Minimum Orbit Intersection Distance with the Earth.} (<0.05 AU) NEOs)
 have tracklets with low values of $\DD$ . 

    \begin{figure}[htp]
    \centering
    \includegraphics[width=\columnwidth]{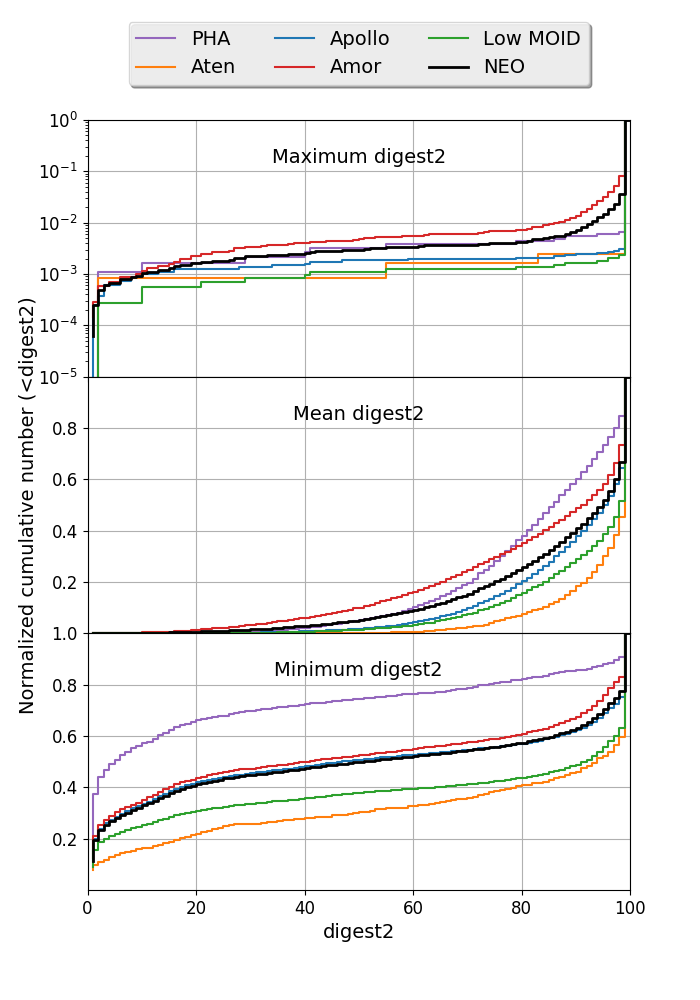}
     \caption{%
     Cumulative histogram of maximum, mean and minimum \D{} score per object for NEOs and their orbital subclasses from $Sim$ data set. Only 0.4\% of NEOs never reach  $D>65$.}
     
        \label{FIG:PVa}
     
    \end{figure}

Figure~\ref{FIG:PVa} shows the cumulative histograms of the minimum, mean and maximum $\DD{}$ per object from the $Sim$ data set, and the differences in $\DD{}$ seen between the different orbit classes.
NEOs can, at times, have low scores: 53\% of NEOs have $\DD{} < 65$ at some point while observable.
But in general, NEOs (and their subclasses) achieve high $\DD{}$ in almost all cases: 
94\% of NEOs reach $\DD{}=100$ (Table~\ref{t:scoresSim}). 
This ratio is highest for objects that come close to the Earth or the Sun ( Aten, Apollo and low-MOID classes). 
The ratio is smallest in the case of Amors, only 87\% of which ever reach $\DD{}=100$, because Amors remain distant and often mimic Main-belt motion when near aphelion. 
\emph{ Only 0.4\% of NEOs never reach $D>65$ while visible. }

\begin{table}[htp]
\small
\begin{center}
\footnotesize
\caption{Fraction of NEOs that achieved a given $\DD{}$ in the $Sim$ data set. }
\begin{tabular}{l|ccc}
\hline
Orbit type   & min $\DD{}>65$ & max $\DD{}<65$  &max $\DD{}=100$\\
&(\%) &(\%) &(\%)\\
\hline
NEO&46.7   & 0.4& 93.9\\
PHA &22.8   & 0.4& 99.2\\
Aten  &  65.9  &  0.2& 99.6\\
Apollo & 46.4   & 0.2& 99.5\\
Amor  & 43.8   & 0.6& 86.4\\
Low-MOID &   59.9  &  0.1& 99.5\\
\hline
\end{tabular}
\label{t:scoresSim}
\end{center}
\end{table}

\begin{figure*}[ht]
\centering
\includegraphics[width=\textwidth]{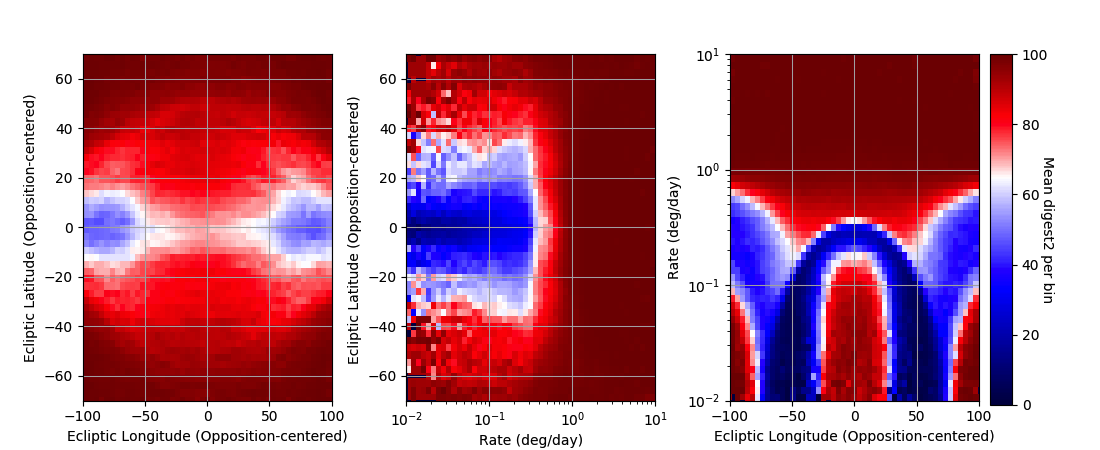}
\caption{\D{} score for all of the NEOs from the $Sim$ data set in opposition-centered ecliptic coordinates per-bin mean \D{}. 
\D{} is generally smaller at low solar elongation. }

%

%
\label{fig:ecliptic_grid}
\end{figure*}

To understand the NEOs with low $\DD{}$, in Figure~\ref{fig:ecliptic_grid} we plot the rate-of-motion and sky-plane location (in ecliptic coordinates centered at opposition ) of each tracklet from the $Sim$ data set.
There are two regions in the morning and evening sky where NEOs tend to have lower $\DD{}$. 
NEOs in these regions are moving very slowly and mimic the Main-belt rate. These regions are coincident with so-called ``sweet-spots'' at low solar elongations, that are the only regions of the sky in which many NEOs become observable \citep[e.g.][]{STOKES03,CS04,Boattini07}. 
There is also an area near opposition where low digest scores are possible for NEOs moving with rates of motion similar to those of MBAs. Interestingly, if the object moves even slower, e.g. below 0.1 degree per day, its score is very high. 
All fast moving objects (rates above $\sim0.8$ degrees/day), and all objects at large observed ecliptic latitudes, have $\DD{}=100$.

\subsection{Accuracy, Precision and the NEOCP}
\label{SECN:RES:ACCURACY}
To understand how accurate $\DD{}$ is when used as an NEO classifier, we can calculate the value of $\DD{}$ for tracklets from a population of known objects (both NEOs and non-NEOs). 
We can then compare $\DD{}$ to an imposed critical value $D_{2,Crit}$, and calculate for the population the rate of ``True Positives''($TP$), ``False Positives'' ($FP$), ``False Negatives'' ($FN$), and ``True Negatives'' ($TN$).
Repeating this gives the values of these quantities as a function of $D_{2,Crit}$. 
We have summarized these quantities (and some additional metrics) in Table \ref{t:EM_Def}.

\begin{deluxetable}{l|rr}
  \caption{%
  {\bf Top:} Error matrix for \D{} when used as a classifier, i.e. when the \D{} score, $D_2$, is compared to a critical value $D_{2,Crit}$.
  {\bf Bottom:} Definition of additional useful metrics.
  }
  \label{t:EM_Def}
  \tablehead{ & \colhead{NEO} & \colhead{Non-NEO}}
  \startdata
  $D_2 >= D_{2,Crit}$ & True Positives, $TP$ & False Positives, $FP$ \\
  $D_2 <  D_{2,Crit}$ & False Negatives, $FN$ & True Negatives, $TN$ \\
  \hline
  \hline
\multicolumn{3}{l}{False Positive Rate, $FPR=FP/(FP+TN)$}\\ 
\multicolumn{3}{l}{False Negative Rate, $FNR=FN/(FN+TP)$}\\ 
\multicolumn{3}{l}{Precision, $PRE=TP/(TP+FP)$}\\
\multicolumn{3}{l}{Accuracy,  $ACC=(TP+TN)/(TP+TN+FP+FN)$}\\
  \hline
  \enddata
\end{deluxetable}


\begin{figure}[htp]
\centering
\includegraphics[width=\columnwidth]{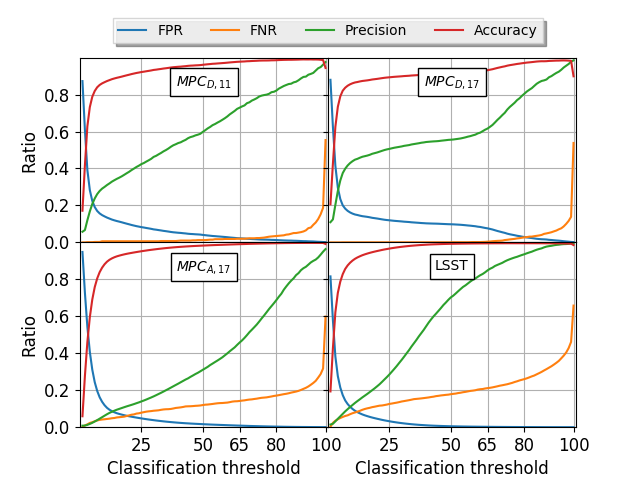}
\caption{
    {\bf Top Left:} $MPC_{D,11}$, All discovery tracklets from 2011;  
    {\bf Top Right:} $MPC_{D,17}$, All discovery tracklets from 2017; 
    {\bf Bottom Left:} $MPC_{A,17}$, All tracklets from January 2017.
    {\bf Bottom Right:} $LSST$, All tracklets from LSST sim.
    Each panel plots 
    the False-Positive-Rate (FPR, blue), 
    the False-Negative-Rate (FNR, orange), 
    the Precision (Green), 
    and the Accuracy (red).
    The precision is a measure of the purity of the tracklets that will end-up on the \NEOCP{}.  
    The ``kink'' that is particularly obvious in the ``Precision'' measure for $MPC_{D,17}$ is driven by the $D_{2,Crit}=65$ NEOCP threshold.
  }
\label{f:PREC}
\end{figure}      


As described in Section \ref{s:Classifier}, unknown tracklets with $\DD{}>D_{2,Crit}$ are submitted to the NEOCP, increasing the likelihood of follow-up observations being obtained and hence of the object being confirmed as real. 
The precision, $PRE$, is a measure of the purity of the tracklets that will end-up on the NEOCP (i.e. what fraction of the tracklets on the NEOCP will actually be NEOs).

In Figure \ref{f:PREC} we plot the \emph{False Positive Rate}, \emph{False Negative Rate}, \emph{Precision}, and \emph{Accuracy} as functions of the critical \D{} score, $D_{2,Crit}$. 
We do this  for both real ($MPC_{D,11}$, $MPC_{D,17}$ and $MPC_{A,17}$) and synthetic ($LSST$) data sets. 
These plots show that the \emph{discovery} data sets ($MPC_{D,11}$ and $MPC_{D,17}$) both display ``kinks'' in the various performance metrics at the values of $D_{2,Crit}$ used at the time of submission.
Above the threshold, most NEOs are followed-up and orbits computed. However, NEOs below the threshold at the time of first observation are less likely to be immediately followed-up.
In contrast, no selection effects are imposed on either the $MPC_{A,17}$ or $LSST$ data sets, hence no kinks appear.

We find that the false negative rate is the largest in the  $MPC_{A,17}$ and $LSST$ data sets, essentially because tracklets with $\DD{}<\DD{}_{,crit}$ are absent from the other ($MPC_{D,11} \& MPC_{D,17}$) data-sets.

We note that the value of $D_{2,Crit}=65$ was chosen with the aim of generating approximately half of the objects on the NEOCP being NEOs. In 2010, the $D_{2,Crit}$ was lowered to $D_{2,Crit}=50$ based on the request of the community and availability of more follow-up and discovery assets. After a year, due to the lack of follow-up of the low-scoring NEO candidates, MPC increased $D_{2,Crit}$ back to $D_{2,Crit}=65$ in mid-2012.
It is clear from both the curves in Figure \ref{f:PREC} and by \citet{Veres2018} that this choice was successful in generating the approximate desired level of precision. 
If members of the community wish to advocate for changes to the adopted value of $D_{2,Crit}=65$, they should weigh-up the pros and cons of such a change: Increasing the value of $D_{2,Crit}$ will cause the false positive rate to drop (cutting down on the number of non-NEOs ``unnecessarily'' followed-up), but the false negative rate will rise (hence losing NEOs).

\subsection{Nuances of Practical \D{} Usage}
\label{s:practical}

There are a number of practical details relating to the algorithmic design of \D{} that should be considered when using the \D{} code in practice.

\subsubsection{Effects of Random-Choice of $\alpha$}
\label{s:random}
%
\begin{figure*}[ht]
\begin{minipage}[b]{\textwidth}
\centering
  \includegraphics[trim = 0mm 0mm 0mm 0mm, clip, angle=0, width=0.33\textwidth]{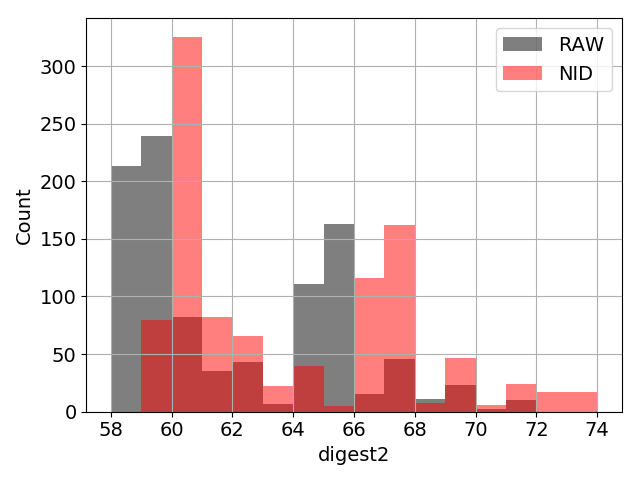}
  \includegraphics[trim = 0mm 0mm 0mm 0mm, clip, angle=0, width=0.33\textwidth]{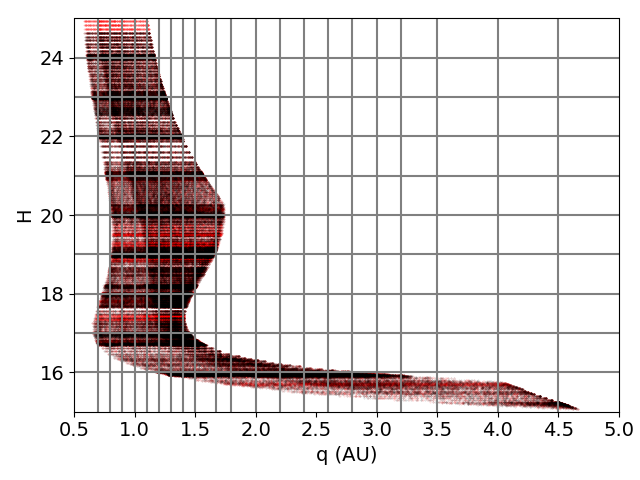}
  \includegraphics[trim = 0mm 0mm 0mm 0mm, clip, angle=0, width=0.33\textwidth]{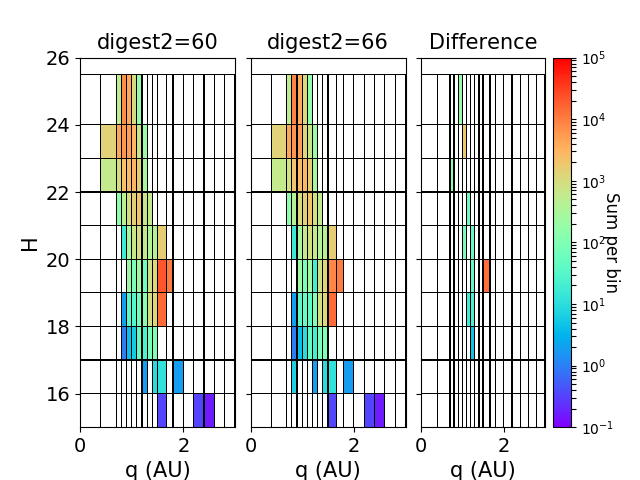}
  \caption{{\bf Left:} One thousand runs of digest2 on a single Pan-STARRS1 tracklet P10Gj15, resulting in a range of 'raw' and 'nid' NEO scores due to the random selection of $\alpha$. This is a particularly extreme example of the variation that is seen. 
  {\bf Center:} Variant orbits generated in two independent runs of \D{} (which resulted in different $\DD{}$ values: Red $\DD{}=60$ and black $\DD{}=66$) plotted in perihelion distance (q) and absolute magnitude (H).
  {\bf Right:} Population bins populated by the variant orbits from the center panel. The different runs hit different bins, causing skews in the calculated score.
  }
 \label{fig:two_scores_alpha}
\end{minipage}
\end{figure*}

The \D{} code uses a pseudo-Monte Carlo method to generate a range of variant orbits (see Appendix~\ref{app:admiss}). 
By default, the pseudo random number generator is seeded randomly. 
Therefore, even though the admissible region is the same, the sampling of available orbits could hit different population bins during an independent instance of the program. 
We demonstrate an extreme example of the $\DD{}$ variation using a real NEO candidate (\emph{P10Gj15}) that was observed on January 17, 2018 as a 3-detection tracklet. The tracklet had $\DD{}=66$ and therefore it was posted to the NEOCP. 
When we re-ran \D{} 1000-times on the tracket, the distribution of $\DD{}$ in the left-panel of Figure~\ref{fig:two_scores_alpha} shows that on many occasions the $\DD{}$ value was below 65. 

To understand this result, we show two specific runs, one leading to $\DD=66$ and one to $\DD{}=60$ in the center of Figure~\ref{fig:two_scores_alpha}, demonstrating that the sampled orbits inhabit almost identical regions of phase space.
However, counting the bins hit by generated variant orbits in the two runs (illustrated in the right-hand side of Figure~\ref{fig:two_scores_alpha}), we see that different population bins were hit, giving different scores. 
In the case of $P10Gj15$, further follow-up observations allowed the orbit determination of the object that became announced as $2018\,BE1$ - a Hungaria-class asteroid.

We repeated this analysis for 1000 randomly selected tracklets from the $Granvik$ NEO data set, running the $\D{}$ code 100-times for each.
We found that 82\% of the tracklets had no change in the integer value of $\DD{}$.
In particular, tracklets with $\DD{}\sim0$ or $\DD{}\sim100$ had almost no variation in $\DD{}$. 
Only a small fraction of tracklets, 0.7\%, had a variation $\Delta\DD{}>3$. 
Most of the varying $\DD{}$-tracklets were in the range $40<\DD<60$, the region where both NEO and Main-Belt populations seem to be probable for a given velocity vector.

The small variation is caused by the binned-population model and the algorithmic design. Future versions of the code should reduce/eliminate this effect. 

We emphasize that if the user wishes to suppress these variations, the keyword \emph{repeatable} can be added to the configuration file. 
The random angle $\alpha$ is then reseeded with a constant value for each tracklet, yielding repeatable scores in independent runs.

\subsubsection{Effects of Observational Uncertainty}
\label{s:obs err}

As illustrated in Section \ref{s:Dithering}, ``dithered'' tracklets are constructed from two end-points or two generated points of the tracklet. The dithering itself is governed by the expected astrometric uncertainty. The default astrometric uncertainty used by \D{} is 1.0 arcsecond. 
However, in our work and at the MPC, astrometric uncertainties are assigned in the $\D{}$ configuration file, using the long-term  average for each observatory code (Table~\ref{t:tabUnc}, Appendix~\ref{app:tables}). 
%
\begin{figure}[ht]
\begin{minipage}[b]{\columnwidth}
\centering
  \includegraphics[trim = 0mm 0mm 0mm 0mm, clip, angle=0, width=0.99\textwidth]{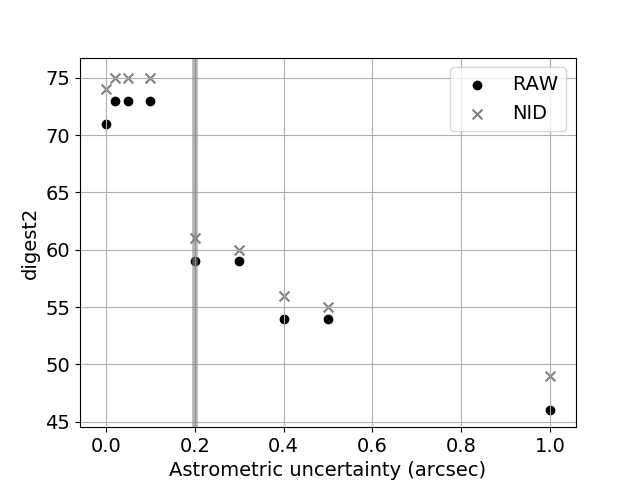}
  \caption{ 
  $\D{}$ score computed for a Pan-STARRS1 tracklet P10Gj15, assuming different astrometric uncertainty for each run. 
  The vertical line shows the assumed Pan-STARRS1 uncertainty.
  The variation in score is driven by the same effect demonstrated in Figure \ref{fig:two_scores_alpha}: varying the uncertainty changes the population bins used to calculate the tracklet's score.  
 }
    \label{fig:uncObj}
\end{minipage}
\end{figure}
%
To demonstrate the effect of changing the assumed uncertainty on a single tracklet, we again chose the Pan-STARRS tracklet $P10Gj15$. Figure~\ref{fig:uncObj} shows how $\DD{}$ depends on the assumed uncertainty. 
The variation in score is driven by the same effect demonstrated in Section \ref{s:random} (Figure \ref{fig:two_scores_alpha}): varying the uncertainty changes the population bins included in the tracklet score calculation. 
In the example illustrated, when the astrometric uncertainty is overestimated, $P10Gj15$ $\DD{}$ leads to more variant orbits that hit the Main-Belt bins.

To study the significance of the assumed astrometric uncertainties for a large data set, we selected 30,000 NEOs from the $Granvik$ data set, generated two-detection tracklets with 0.2" astrometric errors and computed $\DD{}$ using three different assumed astrometric uncertainties: underestimated (0.05"), nominal (0.2") and overestimated (1.0").
Note, that the analysis was undertaken with the \D{} keyword ``repeatable'' set to avoid variation due to angle $\alpha$ randomness. 

When the astrometric uncertainty is overestimated by 0.8" (underestimated by 0.15"), the value of $\DD{}$ varies by an amount $\Delta\DD{}\approx-1$ ( $\Delta\DD{}\approx+0.2$). 
This effect is particularly obvious for intermediate values of $\DD{}$ (20 to 80), where overestimation (underestimation) lead to  $\Delta\DD{}\approx-5$ ( $\Delta\DD{}\approx+1.0$). 
$\DD{}$ values near zero and 100 do not show any variation. 
Overall there was no change for 90\% of the "underestimated" and 78\% of the "overestimated" tracklets. 

\subsubsection{Great Circle Departure}
\label{s:GCR}

\begin{figure*}[ht]
\begin{minipage}[b]{\textwidth}
\centering
  \includegraphics[trim = 0mm 0mm 0mm 0mm, clip, angle=0, width=0.33\textwidth]{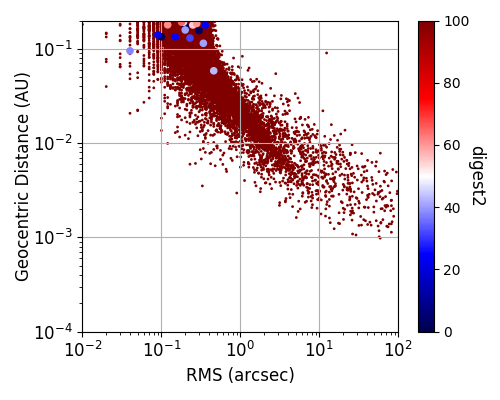}
  \includegraphics[trim = 0mm 0mm 0mm 0mm, clip, angle=0, width=0.33\textwidth]{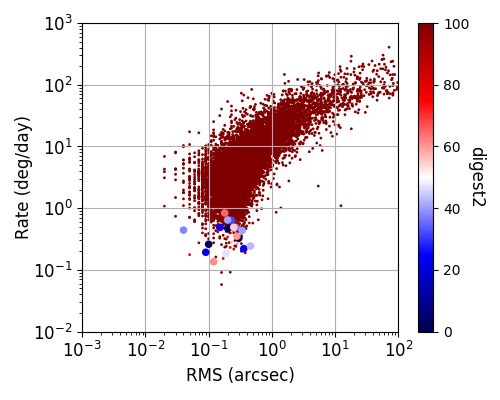}
  \includegraphics[trim = 0mm 0mm 0mm 0mm, clip, angle=0, width=0.33\textwidth]{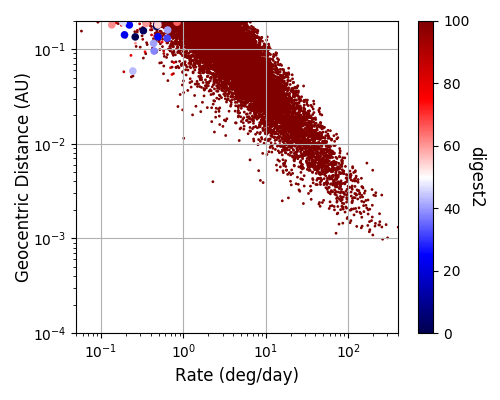}
  \caption{
    $\DD{}$ score of close-approaching NEOs derived for 4-detection tracklets. Correlations between great-circle-residual $rms$, geocentric distance  and rate of motion are depicted. 
    Nearby objects exhibit large $rms$ and high rates-of-motion, but also always have $\DD{}=100$. 
    Large points are used to emphasize those NEOs with low $\DD{}$. There are \emph{no} fast-moving, nearby NEOs or NEOs with non-linear motion, that have \emph{low} digest score. }
 
 \label{f:rms1}
\end{minipage}
\end{figure*}

As described in Section~\ref{s:Dithering}, the \D{} algorithm uses a pair of synthetic detections derived from the set of observed positions. 
In a typical short arc tracklet spanning $\sim 1$hour, the motion is mostly linear. 
However, when the object is close to the Earth, the effect of diurnal parallax for the topocentric observer can cause discernible curvature that is seen as a deviation from the great circle, even within one hour. 
Using the keyword "rms" will cause the \D{} code to output the root-mean-square of the great circle residuals for the positions reported in a tracklet (if the tracklet has $\ge3$ detections). 
If the $rms$ score is within the expected astrometric uncertainties, then, linear motion is a good approximation for the \D{} method. 
A large $rms$ can be due to either bad astrometry, or to diurnal parallax.

To test the effect of great circle deviations we selected close approach events from the CNEOS web site\footnote{\url{https://cneos.jpl.nasa.gov/ca/}} for asteroids that approached the Earth within $0.2\au$ during 2006-2018. 
We generated 16,800 events for 11,828 NEOs. 
We generated 4-detection tracklets spanning 1 hour for each event.
We selected the Pan-STARRS1 $F51$ observatory code as the topocentric position and ``smeared'' the generated astrometry by an expected astrometric uncertainty of 0.2 arc-second. 
Only tracklets $>60$ degrees from the Sun were considered. 

Figure~\ref{f:rms1} illustrates the rate of motion, the $rms$ of the great-circle-residuals, and the geocentric distance of each tracklet.
As expected, the $rms$ increases with decreasing geocentric distance. 
In addition, the correlation between rate of motion and $rms$, as well as between rate of motion and geocentric distance is obvious. 
Low scoring NEOs had $rms<0.2"$, their rate of motion was consistent with the rate of Main-belt asteroids and were distant at the same time.

\subsection{Earth quasi-satellites, co-orbitals and Trojans}
\label{coorbitals}

\begin{figure*}[htp]
\centering
\includegraphics[width=0.9\columnwidth]{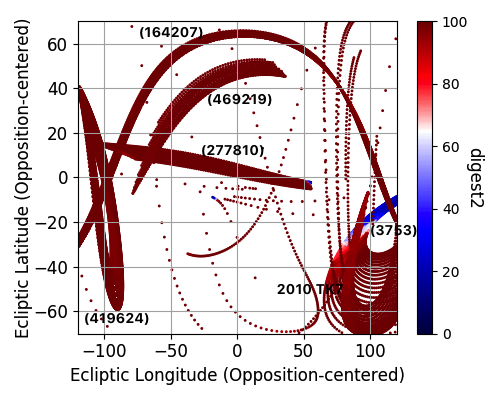}
\caption{\D{} score for known Earth quasi-satellites and their sky-plane motion in ecliptic coordinates centered at the opposition. \D{} score is mostly large. Low \D{} scores corresponds to the ``sweet-spots'' areas. Pattern of the horseshoe orbits is apparent at low solar elongations.}
\label{fig:coorbitals}
\end{figure*}

Although the Moon is the only known natural satellite of the Earth, temporarily captured NEOs, NEOs in 1:1 mean-motion resonance with the Earth, Earth Trojans or proposed minimoons\citep{2012Icar..218..262G,minimoons} can follow geocentric orbits for days or months. (469219) 2016 HO3 is the closest and most stable Earth's quasi-satellite \citep{2016MNRAS.462.3441D}. The only Earth Trojan discovered so far is 2010 TK7\citep{2011Natur.475..481C,2012ApJ...760L..12M}. We selected a small group of known quasi-satellites or co-orbitals\footnote{(3753)  Cruithne, (164207), (277810), (419624), (469219), 2002 AA29, 2006 RH120, 2010 TK7}, generated daily ephemerides from 2000 until 2020, created 20-minute tracklets and simulated sky-plane visibility down to +22.5 $V-band$ magnitude. Their sky-plane motion and \D{} score are shown in Figure~\ref{fig:coorbitals}. The analyzed co-orbitals typically display large valuess of $\DD{}$, unless they are in the low-scoring ``sweet-spots''.  The overall distribution of $\D$ scores is similar to $MPC_{A,17}$ and  $Granvik$ from Figure~\ref{f:NEO_d2} in Section~\ref{s:NEOs}. The only known Earth Trojan 2010 TK7 has a $\D{}$ value oscillating between 93 and 100, and  $40\%$ of its tracklets had $DD{}=100$.

\subsection{NEOs at low Solar elongations}
\label{nearSun}

\begin{figure*}[htp]
\centering
\includegraphics[width=\textwidth]{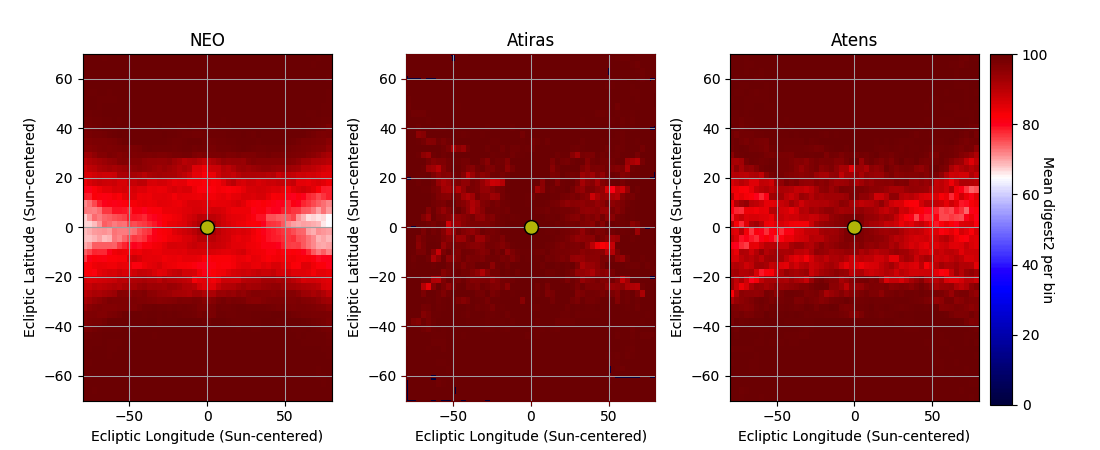}
\caption{\D{} score for all of the NEOs, Atiras and Atens from the $Sim$ data set in Sun-centered ecliptic coordinates per-bin mean \D{}. 
\D{} is generally large at low solar elongation. Yellow point in the center represents the Sun.}
\label{fig:closeToSun}
\end{figure*}

Most of our analyses were focused on NEOs observed in favorable locations of the sky within 100 degrees of the opposition (e.g. Figure~\ref{fig:ecliptic_grid}). In Section~\ref{s:NEOs_details} we identified regions near the ecliptic roughly 100 to 60 degrees from the opposition where the mean \D{} score is relatively low. However, some orbit types, such as Atens or Atiras, rarely or never cross the opposition region.  Searches close to the Sun are required to find these objects. Ground-based observations near the Sun are limited by large airmass and associated target brightness losses, as well as limited observing time at dawn or dusk. Low Solar elongation observations are ideal for space-based observatories. Figure~\ref{fig:closeToSun} shows the mean \D{} score for the $Sim$ data set in the Sun-centered ecliptic coordinates. Interestingly, closer to the Sun \D{} is almost always above 65 and Atiras have particularly very large $\DD{}$ values. We demonstrated that $\D{}$ score can identify NEOs in low-solar elongations.

\subsection{Interstellar Objects and \D{}}

The interstellar object (ISO) 1I/'Oumuamua \citep{2017MPEC....U..181B} was discovered by the Pan-STARRS NEO survey and its first tracklet was reported to the NEOCP as a $\DD{}=100$ tracklet. 
Subsequent follow-up quickly revealed its hyperbolic orbit. However, without a large NEO $\D{}$ score, it would never have appeared on the NEOCP or been followed-up.
As mentioned earlier, one of the key constraints of \D{} are orbits bound to the Sun.
Previous work \citep{Engelhardt} showed that the NEO \D{} score could be used to identify ISOs, demonstrating that 2/3 of ISOs in a simulated Pan-STARRS survey achieved $\DD{}>90$ at some time while visible. 

\begin{figure}[htp]
    \centering
    \includegraphics[width=\columnwidth]{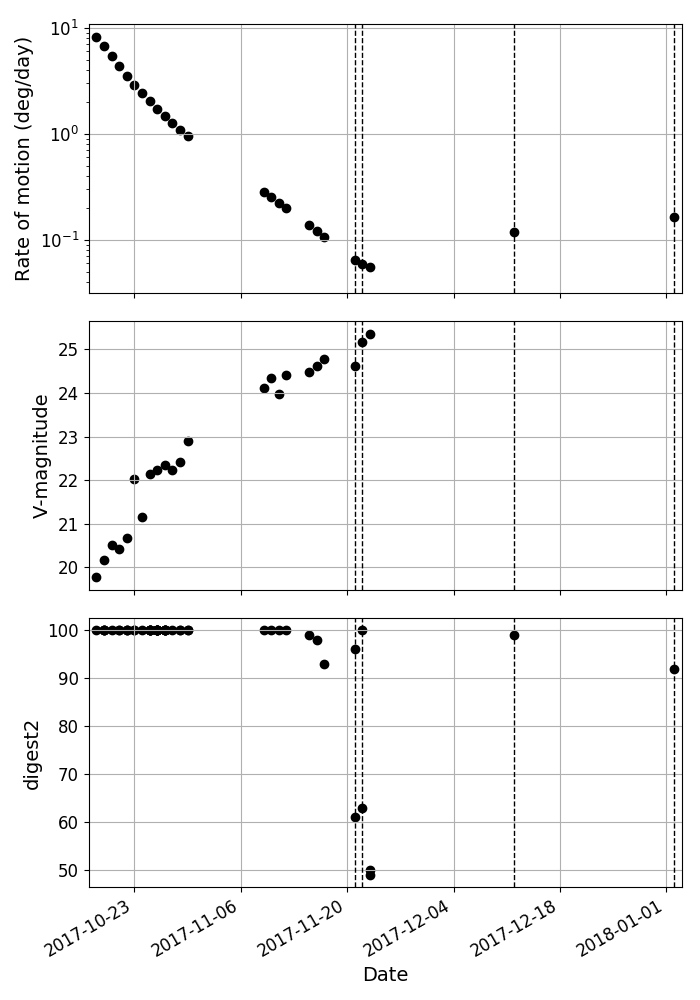}
     \caption{%
     \D{} score, mean V-band magnitude and rate of motion of 1I/'Oumuamua tracklets. Dashed vertical lines display epochs when object was observed by Hubble Space Telescope.
    1I/'Oumuamua has a high \D{} score for the entire time it was visible to surveys ($V\lsim22$). 
        \label{FIG:oumuamua}
     }
    \end{figure}

Figure~\ref{FIG:oumuamua} shows the $\DD{}$, rate of motion and magnitude of all reported 1I/'Oumuamua tracklets. 
From the time of discovery to November 12, 2017, 1I/'Oumuamua had $\DD{}=100$ at all times, and in addition, 1I/'Oumuamua had $\DD{}=100$ for the entire time it was visible to the main NEO surveys ($V\lsim22$).

\section{Discussion}
\label{s:DISC}

We have described the algorithm underlying the \D{} code, as well as its usage to calculate the score of a short-arc tracklet with respect to a given orbit class. 
We focused on the \emph{NEO} \D{} score, $\DD{}$, as used by the Minor Planet Center (MPC) to classify the likelihood that submitted sets of detections are NEOs. We showed that $\DD{}$ changes for each object as it moves across the sky and its brightness and rate of motion vary. 
For NEOs, the maximum $\DD{}$ reaches the extreme value of $\DD{}=100$ for more than 93\% of NEOs, and over $99\%$ of NEOs have $D_{2,crit}>65$ at some point while visible.  
Thus, the vast majority of NEOs have sufficiently high $\DD{}$ to allow them to be posted to the NEOCP and be rapidly followed-up by observers around the world.

While the \D{} code identifies short-arc NEOs correctly in the vast majority of cases, there are a small number of cases when $\DD{}$ is below the critical threshold, currently set to $D_{2,crit}>65$. 
We identify the two locations on the sky, on the ecliptic and at low solar elongations, where the mean $\DD{}$ is significantly lower than elsewhere on the sky. This is due to the observing geometry and rates of motion of the objects, that cause them to be confused with Main-Belt objects.

We also demonstrate that without the "repeatable" parameter the \D{} code can generate varying output for the same tracklet due to its random selection of variant orbits. However, the variation in $\DD{}$ is small and does not affect scores near extremes ($\DD{}=0$ and $\DD{}=100$). 
The MPC uses \D{} without the "repeatable" parameter to avoid bias.

\subsection{Future directions}
\label{s:DISC:Future}

As we move into an era of large surveys that can obtain huge volumes of high-precision observations and utilize highly-precise stellar catalogues, it is clear that the \D{} code must evolve. 
Future developments to the code will include:

\begin{itemize}
\setlength\itemsep{0.01em}
\item It will be possible to characterize observational uncertainties on a per-observation basis, as provided for in the ADES format\footnote{\url{https://minorplanetcenter.net/iau/info/ADES.html}}.
\item A python wrapper and API to provide simplified usage. 
\item A new NEO population model, such as the high-fidelity \citet{Granvik18} model, will be required. Catalog of known orbits used for unknown population reduction would be generated more frequently.
\item The discrete population bins (in $q,e,i, H$) need to be replaced with a smooth 4-dimensional function, perhaps also incorporating additional orbital elements to allow improved discrimination of (e.g.) Hildas and Jupiter Trojans.
\item Replacing the synthetic two-detection approach with a robust statistical treatment of all submitted detections, allowing for a significantly improved handling of NEOs that exhibit non-linear motion during close-approach.

\item Further tests and improvements of the code are needed for space-based NEO surveys, such the proposed NEOCam mission \citep{2015AJ....149..172M}.

\item Replacing the $astorb$ known orbit catalog and its orbit quality metric \citep{1993Icar..104..255M} with the MPC's orbit catalog\footnote{\url{https://minorplanetcenter.net/iau/MPCORB.html}} and orbit uncertainty parameter\footnote{\url{https://minorplanetcenter.net/iau/info/UValue.html}}.

\end{itemize}

We emphasize that \D{} constitutes an evolving code-base and that we welcome reports from the community of any bugs found, or of suggestions for future directions\footnote{At the time of submission, the \D{} team are currently diagnosing the cause of a rare output-formatting error that appears to afflict approximately 1 in every 50,000 evaluation attempts when running \D{} in a multi-core environment. We anticipate that the solution of this issue will lead to a future version release.}.

\acknowledgments
We are grateful to Rob McNaught for his insights regarding the history of the \PG code, and for his many clarifications regarding the historical development of ranging.

MJH and MJP gratefully acknowledge NASA grants NNX12AE89G, NNX16AD69G, and NNX17AG87G, as well as support from the Smithsonian 2015-2017 Scholarly Studies programs.
DJA acknowledges the N. Ireland Dept.\ for Communities support
at Armagh.

We are saddened to report that during the course of the work on this manuscript, lead author Sonia Keys died on 13 August 2018 at the age of 57, after a long struggle with cancer.
Sonia worked as a commercial software developer from 1983 to 2001 and then began working with the Astronomical Society of Kansas City, specializing in the tracking of Near-Earth Objects (NEOs).  
Her skill and reliability with NEO astrometry caught the attention of the Minor Planet Center (MPC), and that led to a position as an astronomer and software developer for the MPC.
During her tenure at the MPC, Sonia was the lead developer of the \D{} software and was the point-of-contact for the many people in the NEO community that utilized it for over a decade.
We deeply regret Sonia's passing, and we hope that we have completed this manuscript in a manner she would have approved of. We hope this publication and note will serve as a small token of the esteem in which we held her. 
Sonia was the conscience of the MPC.  
Her clarity of thought, her willingness to ask difficult questions, and her tenacity will be sorely missed.

\appendix

\section{Variant Orbits}
\label{app:admiss}
%

\begin{figure}[hpt]
\centering
  \includegraphics[width=10cm]{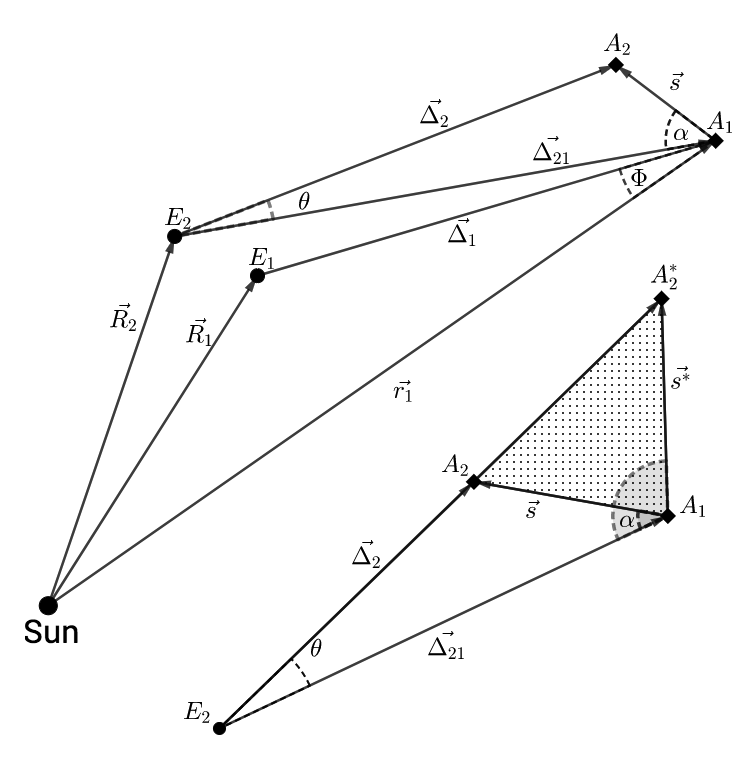}
  \caption{%
    {\bf Top:} Vector algebra for solving state vectors according to Appendix~\ref{app:admiss}. {\bf $E_i$ and $A_i$ are the positions of the Earth and the asteroid at times $t_i$.}
    {\bf Bottom:} Because only the transverse component of the velocity vector is visible to the observer, the true velocity vector could be in the dotted area between
    $A_2$ and $A_{2}^*$, that represents all possible solutions for bound orbits. $\vec{\Delta_{i}}$ represent topocentric and $\vec{r_{i}}$ the heliocentric vectors of the object and $\vec{R_{i}}$ the heliocentric vectors of the observer at time $t_i$. $\vec{s}$ is directed along the velocity vector of the object.
    }
    \label{fig:vec}
\end{figure}

In the following algorithmic description, all line references relate to Algorithm~\ref{alg: essential} of Section \ref{s:COMP}.

The input file for the \D{} program contains astrometry of short-arc tracklets, such as those illustrated in Appendix~\ref{app:Hungaria tracklet}.  
The file can contain one or many tracklets (Line~\ref{ess:load_obs}). 
After the input data are loaded, binned population model is loaded into memory as well (Line~\ref{ess:load_popn}, Section~\ref{s:POPN}). Each tracklet is subsequently treated individually (Line~\ref{ess:read arc}):

Two observations of an arc are insufficient for a determination of its orbit, or when the observation arc is too short. 
However, even very short arcs with few observations can be used to derive a range of possible orbits and assess the probability of the asteroid being in a certain region of the Solar system. Because each tracklet has two or more detections (N), as described in Section~\ref{s:EndPoint}, two detections either selected ($N=2$) or generated ($N>2$).

The two observables, or detections, $A_1$ and $A_2$ of the same object (Line~\ref{ess:two pos}), are defined by $A_{t_i}$=[$\alpha_i, \delta_i, V_i$], 
where 
$\alpha_i$ is the right ascension,
$\delta_i$ the declination, $V_i$ is the apparent magnitude at the time of observation $t_{i}$.

Because the tracklet consists of $N\geq2$ data points with unique photometry, either due to photometric errors or rotation of the object, the mean tracklet magnitude $V_1$ is computed (line~\ref{ess:V mag}) in the Johnson-Cousins V-band photometric system. The transformation to V-band is simple \citep{Veres2018}, the correction is $-0.8$ if the reported band was $B$, or $+0.4$ for other bands, except of $V$, where no correction is applied. If no photometry is provided, $V_1=21$ is assumed as a typical magnitude limit of the current asteroid surveys.

We can easily transform the observation [$\alpha_i, \delta_i$] to its topocentric cartesian unit vector (Figure~\ref{fig:vec}) $\vec{d_i}=[x_i,y_i,z_i]$:

\begin{equation}
(x_i , y_i , z_i) = 
(\cos{\alpha_i} \cos{\delta_i},\,\,\,
\sin{\alpha_i} \cos{\delta_i},\,\,\,\sin{\delta_i}
)
\end{equation}

For further analysis, we will need to transform unit vectors from equatorial to ecliptic coordinate system (See Figure~\ref{fig:vec}):
\begin{equation}
\vec{d_i}= R_{\epsilon}^T \cdot \vec{d_i}
\end{equation}

\noindent where $R_{\epsilon}$ is the rotation matrix defined by obliquity of the ecliptic $\epsilon$:

\begin{equation}
R_{\epsilon}=
\begin{pmatrix}
1 & 0 & 0\\
0 & \cos{\epsilon} & \sin{\epsilon}\\
0 & -\sin{\epsilon} & \cos{\epsilon}\\
\end{pmatrix}
\end{equation}

To derive the position state vector $\vec{r}$, we have to \emph{assume} a topocentric distance $D$ to the asteroid. In line~\ref{ess:dist vec} of Algorithm~\ref{alg: essential}, Sun-observer vectors (labeled $\vec{R_1}, \vec{R_2}$ in Figure~\ref{fig:vec}) in ecliptic Cartesian coordinates are computed by local sidereal time and observer site parallax constants. Observer's topocentric positions are loaded from MPC-managed list of observatory codes\footnote{\url{https://minorplanetcenter.net/iau/info/ObservatoryCodes.html}}.  Also the observer-object unit vectors $\vec{d_1}, \vec{d_2}$ are computed in ecliptic Cartesian coordinates.

Then, the topocentric vector to the first observation, $\vec{\Delta_1}$, will be a scalar multiple of $D$ and the unit vector $\vec{d_1}$. $D$ is one of the values selected recursively (Appendix~\ref{app:bin_search}) between the minimum (0.05\,au) and maximum (100\,au) distances within the allowed admissible region:

\begin{equation}
\vec{\Delta_1}=D\,\vec{d_1}
\end{equation}

Knowing the position vectors of the observer with respect to the Sun at the time of the two observations ($R_1$ and $R_2$) in ecliptic coordinates, we can easily derive the heliocentric state vector $\vec{r_1}$ as:

\begin{equation}
\vec{r_1}=\vec{R_1}+\vec{\Delta_1}
\end{equation}


After the position vectors are derived, the algorithm continues in investigation of the velocity vectors. The observed motion of the asteroid on the celestial sphere is a projection of the actual velocity vector to the tangent plane. While we do not know the angle between the tangent plane and the velocity vector $\vec{v_1}$, we can make progress by computing the possible ranges of angles (Figure~\ref{fig:vec} - bottom) under the assumption that the object is bound to the Sun. 
At a given distance from the Sun, $r_1$, an orbit is bound if its specific orbital energy $\epsilon<0$. 
We will find the angles of the $\vec{v_1}$ for the parabolic orbits from:

\begin{equation}
\epsilon=\frac{v_1^2}{2}-\frac{U}{r_1} = 0
\label{epsilon}
\end{equation}

\noindent where $U=k^2$ is the Gaussian gravitational constant squared\footnote{$k=0.01720209895\,au^{3/2}\,day^{-1}\,(solar\,mass)^{-1/2}$ }. 

\medskip\noindent To find $\vec{v_1}$ one has to solve triangle $E_2, A_1, A_2$ on figure~\ref{fig:vec}. The velocity vector will be derived as $\vec{v_1}=(\vec{\Delta_2}-\vec{\Delta_{21}})/(t_2-t_1)$.

\medskip\noindent The angle $\theta$ (Figure~\ref{fig:vec} - bottom) can be derived by computing the intermediate vector $\vec{\Delta_{21}}$  from vector difference of $\vec{r_1}$ and $\vec{R_2}$ and the dot product of known unit vector $\vec{d_2}$ and derived $\vec{\Delta_{21}}$:

\begin{align}
\begin{split}
\vec{\Delta_{21}}=\vec{r_1}-\vec{R_2}\\
\cos{\theta}=\frac{\vec{\Delta_{21}}\cdot\vec{d_2}}{{|\Delta_{21}| |1|}}
\end{split}
\end{align}

\noindent Then, by using the cosine rule, substituting $v_1=s_1/(t_2-t_1)$ and using equation (\ref{epsilon}), we can derive the length of the vector $\Delta_2$ as follows. Line~\ref{ess:angle lim} of Algorithm~\ref{alg: essential} involves solving a quadratic that gives two values for $\alpha$ corresponding to state vectors $\vec{s}$ and $\vec{s^*}$ that represent parabolic orbits:

\begin{align}
\begin{split}
{\Delta_{21}}^2+{\Delta_{2}}^2-{s_{1}}^2-2{{\Delta_{21}}\Delta_{2}}\cos{\theta}=0\\
{\Delta_{2}^2}-2{{\Delta_{2}}{\Delta_{21}}}\cos{\theta}+\left({\Delta_{21}}^2-2\frac{U{(t_2-t_1)}^2}{{r_{1}}}\right)=0
\end{split}
\label{roots}
\end{align}

\noindent Equation (\ref{roots}) has two roots for $\Delta_2$ after doing simple substitution (A, B, C) solving:

\begin{align}
\begin{split}
(A,B,C) &= (1 , \,\,\, -2\Delta_2\cos{\theta} , \,\,\, {\Delta_{21}}^2-2\frac{U{(t_2-t_1)}^2}{{r_{1}}})
\\
\Delta_2&=\frac{- B \pm  \sqrt{B^2-4AC}}{2 A}
\end{split}
\end{align}

\noindent Two roots of $\Delta_2$ lead to the extreme solutions of angles at which the  vector ${v_1}$ of an asteroid is equal to escape velocity:

\medskip\noindent Based on figure~\ref{fig:vec} and cosine rules and law of sines:

\begin{align}
\begin{split}
s_1&=\sqrt{\Delta_{2}^2+\Delta_{21}^2-2\Delta_{21}\Delta_{2}\cos{\theta}}\\
\cos{\alpha}&=\frac{{s_{1}}^2 + {\Delta_{21}}^2 - {\Delta_{2}}^2}{2{s_1}{\Delta_{21}}}  \\
\sin{\alpha}&= \frac{{\Delta_2} \sin{\theta}}{{s_1}}   \\
\alpha&=\arctan{\left(\frac{\sin{\alpha}}{\cos{\alpha}}\right)}
\end{split}
\end{align}

Having the topocentric ($\vec{r_1}$) and heliocentric ( $\vec{\Delta_1}$) distance, the mean V-band magnitude $V_1$ and geometry ($\alpha$), we are now able to compute the object's absolute magnitude $H$ (Line~\ref{ess:H mag}) of Algorithm~\ref{alg: essential} by equation:

\begin{equation}
H=V_1-5\log(\Delta_1 r_1) +f(\Phi_1, G)
\end{equation}

\noindent where the phase angle $\Phi_1$ is the angle between $\vec{\Delta_1}$ and $\vec{r_1}$ and $G$ is the slope parameter. 
The form of the phase function $f(\Phi,G)$ is based on \citet{1989aste.conf..524B} and $G=0.15$ is assumed. \\

In the next step (Line~\ref{ess:angle loop}) of Algorithm~\ref{alg: essential}, the algorithm picks angles randomly for a selected topocentric distance $D$ in a range between the two extremes of  $\alpha$  that represent a bound elliptical orbit.

With any valid $\alpha$ and $D$, we can derive the velocity vector, again by using the rule of sines (Figure~\ref{fig:vec} - bottom):

\begin{align}
\begin{split}
\Delta_2=\frac{\Delta_{21}\sin{\alpha}}{\sin{(\pi-\alpha-\theta)}}
\end{split}
\end{align}

\noindent The heliocentric velocity vector $\vec{v}$ in ecliptic coordinates is
\begin{equation}
\vec{v}=\frac{{\Delta_2} \vec{d_2} - \vec{\Delta_{21}} }{t_2-t_1}   
\end{equation}

\noindent Having $\vec{r_1}$, $\vec{v}$ at a given epoch, we have the orbit defined through its state vector (Line~\ref{ess:state vec} of Algorithm~\ref{alg: essential})  which is easy to convert into a set of Keplerian elements (Line~\ref{ess:elements} of Algorithm~\ref{alg: essential}).

\begin{figure}[htp]
\centering
  \includegraphics[scale=0.4]{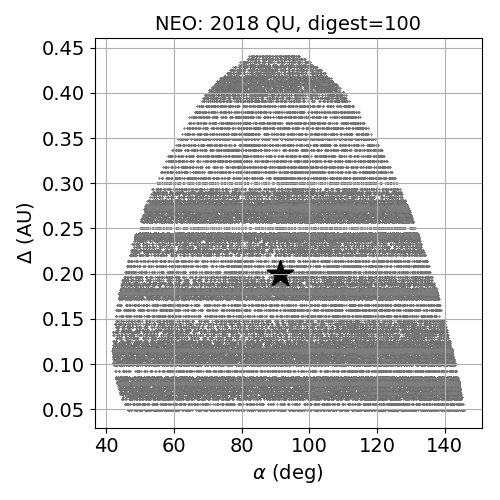}
    \includegraphics[scale=0.4]{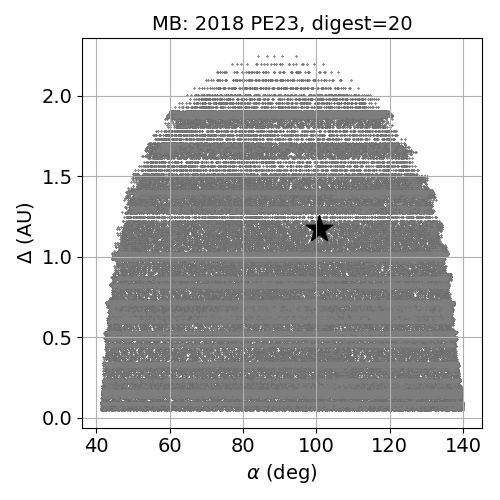}
  \includegraphics[scale=0.4]{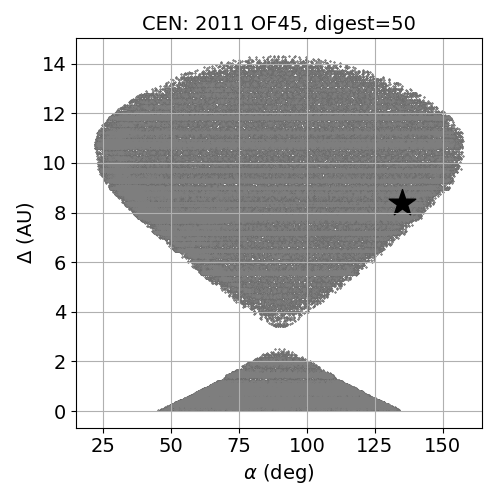}
  \caption{Grid of bound variant orbits generated by \D{} for an NEO (left), Main Belt (center) and Centaur (right) tracklet. The asterisks depict the true position.}
  \label{fig:Trojan}
\end{figure}

\medskip The entire procedure is repeated for a range of assumed distances, $D_j$ that lead to set of \emph{unique} position vectors, $\vec{r_j}$, and velocity vectors, $\vec{v_j}$. Example of variant orbits generated by \D{} in a phase space of $\alpha$ and $\Delta_1$ is shown in Figure~\ref{fig:Trojan}.

\section{\D{} Software}

\subsection{Detailed Historical Evolution}
\label{app:HIST}
In Section \ref{s:HIST} we provided a brief outline of the historical development of the \D{} algorithm. 
In Table \ref{t:Versions} of this appendix, we provide the interested reader with a more detailed, chronologically-ordered description of the key changes made to the \D{} code.

\begin{deluxetable}{l|r|l}[hpt]
\tablecolumns{3} 
\tablewidth{0pc} 
\caption{Significant Changes and Versions of Digest2. Archived (and operational) code for the majority of the versions listed in this table are available in \url{https://bitbucket.org/mpcdev/digest2/src/master/archive/}.  }
    \tablehead{
        \colhead{Date} & 
        \colhead{Version} & 
        \colhead{Description} 
    }
    \label{t:Versions}
    \startdata
        pre-2002 & 0.0223 & Original FORTRAN 77 code (``\texttt{223.f}'') built on \PG{}
        \\
        Aug 2005 & 0.1 & MIT-style license, catalog-reduced model. 
        \\
        Sep 2005 & 0.2 & Bin search algorithm (App. ~\ref{app:bin_search})
        \\
        Feb 2010 & 0.3 & Pan-STARRS S3M model, multiple orbit classes, improved scoring (\S~\ref{s:POPN}).  Parallelized across multiple CPU cores. 
        \\
        Feb 2010 & 0.4 & Supported space-based observations.  
        \\
        Feb 2010 & 0.5 & Great circle RMS output. Stand-alone GCR utility. 
        \\
        Mar 2010 & 0.6 & Improved stand-alone GCR utility.     \\
        Nov 2010 & 0.7 & Orbit classes H18, H22, bug fixes. 
        \\
        Apr 2011 & 0.8 & More flexible output, new config options. Removed stand-alone GCR utility, bug fixes.  
        \\
        Apr 2011 & 0.9 & Bug fixes. 
        \\
        Apr 2011 & 0.10 & Bug fixes. 
        \\
        May 2011 & 0.11 & Endpoint synthesis (\S~\ref{s:EndPoint}).  Observational Error (\S~\ref{s:obs err}).  \\
        May 2012 & 0.12 & Bug fixes. 
        \\
        Jun 2012 & 0.13 & Bug fixes. 
        \\
        Jan 2013 & 0.14 & Bug fixes. 
        \\
        Feb 2014 & 0.15 & Allow more obscodes. 
         \\
        Feb 2015 & 0.16 & Command line option to limit parallelism. 
        \\
        Jun 2015 & 0.17 & CSV model, bug fixes. 
        \\
        Jun 2015 & 0.18 & Minor usability improvements.  Bug fixes. 
        \\
        Aug 2017 & 0.19 & Bug fixes, including a serious uninitialized data problem. \\
    \enddata
\end{deluxetable}

\subsection{Software Availability and Usage}
\label{app:software}
The \D{} source code and the documentation is freely available at \url{https://bitbucket.org/mpcdev/digest2/overview}. 
The download section contains a zip archive of \D{} and the population model tar achive \emph{d2model}.

The synthetic solar system model binning and reduction source code \emph{MUK} is freely available at \url{https://bitbucket.org/mpcdev/d2model/src/master/muk.c}.

The \D{} code can be executed directly after its compilation using the supplied input observation file (sample.obs) and configuration file (MPC.config) as follows:
\begin{verbatim}
digest2 -c MPC.config sample.obs
\end{verbatim}

Further detailed instructions for the operation of \D{} can be found in \url{https://bitbucket.org/mpcdev/digest2/src/master/OPERATION.md}.

\section{Binary Search Algorithm for Population Bins}
\label{app:bin_search}

The initial algorithm (version 0.1) for constructing class scores was as in Algorithm  \ref{alg 0.1}.
\begin{algorithm}
\caption{\D{} scoring version 0.1}
\DontPrintSemicolon
\label{alg 0.1}
    \ldots\;
    \For{\upshape distance $D \gets$ .025 AU to 7 by .025}{ \label{01:dist loop}
        \For{\upshape $\alpha \gets \alpha_1$ to $\alpha_2$ by $2\degr$}{ \label{01:angle loop}
            Look-up $BinPopn(a,e,i,H)$\; \label{01:pop}
            \eIf{$Interesting(a,e,i)$}{ \label{01:int}
                $neo \pluseq BinPopn$\; \label{01:acc neo}
            }{
                $mb \pluseq BinPopn$\; \label{01:acc mb}
            }
        }
    }
    score $\gets 100 * neo / (neo + mb)$\; \label{01:score}
\end{algorithm}

On line~\ref{01:int} the function $Interesting$ tests if candidate elements are in the class of interest. 
The single class of interest defined in \D{} versions-0.1 and -0.2 corresponded to the Minor Planet Center's definition of ``interesting'' which included not just NEOs ($q < 1.3$) but also high eccentricity ($e > 0.5$) and high inclination ($i > 40$) orbits which would also lead to a discovery MPEC.
The current code uses the classes described in Table \ref{t:classes} of Appendix \ref{app:tables}.

Version-0.2 contained an improved score formula.
The goal of the new score formula was to change the computation of the variables neo and mb (Algorithm~\ref{alg 0.1} lines \ref{01:acc neo} and \ref{01:acc mb}) to be more like,
\begin{equation}
\begin{aligned}
neo &= \text{ sum of bin populations over all represented NEO bins}\\
mb  &= \text{ sum of bin populations over all represented non-NEO bins},
\end{aligned}
\label{e:binary1}
\end{equation}
where a best effort is made to locate all the population bins represented by the input arc.
The reasons for this change ultimately stem from a desire to 
(a) avoid double-counting population bins, and 
(b) avoid omitting population bins due to the use of overly large distance and/or angle step-sizes in Algorithm~\ref{alg 0.1}.

Per-class binning enables enhancements to the bin tagging and scoring algorithms.  
In the current version of \D{}, separate bin tags are accumulated per class.  
Further, for each class, two sets of tags are accumulated per class.  An in-class tag is set for a bin when a candidate orbit meets the class definition.  A separate out-of-class tag is set otherwise.
Two population sums are then computed for each class, a sum of bin populations with in-class tags and sum of bin populations with out-of-class tags.  The first represents the sum of class bin populations where a candidate orbit of class was found.  The second represents the sum of out-of-class bin populations where a candidate orbit out of class was found.

Computations at a single distance follow algorithm~\ref{alg: essential} steps~\ref{ess:dist vec} through \ref{ess:H mag}, but then call a recursive function to subdivide the angle space between $\alpha_1$ and $\alpha_2$.

At each call, a state vector and then orbital elements are computed, but then rather than accessing the indicated bin population, the corresponding tag is simply checked (in the distance-specific tags). 
If the tag was already set, subdivision ends and the recursive function returns.  If the tag was not set, then the angle space is subdivided further and the recursive function is called for each subdivision.

The distance, $D$, is subdivided similarly with a separate recursive function. On each call, computations are performed for a single distance as just described, then the distance-specific tags are checked.  If all bins tagged at that distance had previously been tagged in the overall result, the recursive function simply returns.  
If new bins were tagged, they are merged into the overall result, the distance space is subdivided further, and the recursive (distance) function is called for each subdivision.
After all subdivision functions return, the formulas above for $neo$ and $mb$ are computed as described in Algorithm~\ref{tag acc}.

 \begin{algorithm}
 \caption{Tag accumulation}
 \label{tag acc}
 \SetKwFor{ForEach}{for each}{do}{}
 \ForEach{\upshape tag of overall tags}{
    \If{\upshape tag set}{
         \If{\upshape interesting bin}{
            $neo \gets neo + BinPop(a,e,i,H)$
        }
        \If{\upshape non-interesting bin}{
            $mb \gets mb + BinPop(a,e,i,H)$
        }
    }
 }
 \end{algorithm}

\section{Tabulated Values and Settings for \D{}}
\label{app:tables}

In Table \ref{t:tabUnc} we list the assumed astrometric uncertainties adopted in the latest version of the \D{} code.  
The values are set in the file {\sc MPC.config} and can be altered by the user. 

\begin{table}
\small
\begin{center}
\caption{Astrometric uncertainties currently used by \D{} by the MPC and \D{} code, based on long-term statistics\footnote{\url{https://minorplanetcenter.net/iau/special/residuals.txt}}.}
\tabcolsep=0.11cm
\begin{tabular}{c|c||c|c}
\hline
observatory  & astrometric &observatory  & astrometric \\
code & uncertainty & code & uncertainty\\
\hline
106&0.4"&D29&0.5"\\
291&0.4"&E12&0.5"\\
568&0.1"&F51&0.2"\\
691&0.4"&F52&0.2"\\
703&0.7"&G96&0.3"\\
704&0.7"&H15&0.5"\\\
A50&0.5"&J75&0.4"\\
C51&0.7&other& 1.0"\\
\hline
\end{tabular}
\label{t:tabUnc}
\end{center}
\end{table}

\D{} is configurable with keywords that allow control of the output and setup of some input variables. 
The keywords are set in the same file (MPC.config) as the astrometric uncertainties (Table~\ref{t:tabUnc}). 
The available keywords are described in Table \ref{t:keywords}.

\begin{table}[htp]
\small
\begin{center}
\footnotesize
\caption{Keywords available in \D{}.}
\begin{tabular}{l|l}
\hline
keyword & meaning\\
\hline
    headings  & show heading\\
    noheadings & hide heading\\
    rms & show residual RMS from linear motion along a great circle in arc-seconds\\
    norms & do not show RMS \\
    raw & show raw \D{} (Section~\ref{s:POPN:Full})\\
    noid & show noid \D{} (Section~\ref{s:POPN:Unknown})\\
    repeatable & random seed is defined and always the same (Section~\ref{s:random})\\
    random & pseudo-random generation of $\alpha$ and resulting \D{}  (Section~\ref{s:random})\\
    obserr & fixed astrometric uncertainty set to a value in arc seconds\\
    poss & show other non-zero resulting scores in addition to specified classes \D{}\\
    any orbit class(es) defined in Table~\ref{t:classes} &e.g. NEO\\
    user defined astrometric uncertainty (arc seconds) per MPC observatory code&e.g. obserrF51=0.2\\
    \hline
    \multicolumn{2}{l}{default keywords: noid, rms, headings, N22, N18, NEO, Int, obserr=1.0}\\
    \hline
\end{tabular}
\label{t:keywords}
\end{center}
\end{table}

The orbit classes used by \D{} are defined by the orbital elements and absolute magnitudes listed in Table \ref{t:classes}.
\begin{table}[htp]
\small
\begin{center}
\footnotesize
\caption{Definitions of orbital classes in \D{}.}
\begin{tabular}{l|l|l}
\hline
orbit class & abbreviation & definition in keplerian  elements and absolute magnitude\\
\hline
MPC Interesting & Int & q < 1.3 || e >= 0.5 || i >= 40 || Q > 10\\
Near-Earth Object & NEO & q < 1.3 \\
Large Near-Earth Object & N18 & q  < 1.3 \& H<18.5\\
Intermediate-size Near-Earth Objects & N22 & q < 1.3 \& H<22.5\\
Mars-Crossers&MC & q < 1.67 \& q >= 1.3 \& Q > 1.58\\
Hungarias & Hun &a < 2 \& a > 1.78  \& e < 0.18 \& i > 16 \& i < 34\\
Phocaeas &Pho & a < 2.45 \& a > 2.2 \& q > 1.5 \& i > 20 \& i < 27\\
Inner Main Belt & MB1 & q > 1.67 \& a < 2.5 \& a > 2.1 \& i < ((a - 2.1) / 0.4) * 10 + 7\\
Pallas family & Pal & a < 2.8 \& a > 2.5 \& e < 0.35 \& i > 24 \& i < 37\\
Hansas &Han&  a < 2.72 \& a > 2.55 \& e < 0.25 \& i > 20 \& i < 23.5\\
Central Main Belt &MB2 &a < 2.8 \& a > 2.5 \& e < 0.45 \& i < 20\\
Outer Main Belt & MB3 & e < 0.4 \& a > 2.8 \& a < 3.25 \& i < ((a - 2.8) / 0.45) * 16 + 20\\
Hildas & Hil & a > 3.9 \& a < 4.02 \& i < 18 \& e < 0.4\\
Jupiter Trojans & JTr & a > 5.05 \& a < 5.35 \& e < 0.22 \& i < 38\\
Jupiter Family Comets & JFC & q > 1.3 \& $T_{J} > 2$ \& $T_{J} < 3$\\
\hline
 \multicolumn{3}{l}{$T_{J}$ - Tisserand parameter with respect to Jupiter, a - semimajor axis (AU), e - eccentricity, i - inclination (deg)}\\
  \multicolumn{3}{l}{Q - aphelion distance (AU), q - perihelion distance (AU), H - absolute magnitude}\\
\hline
\end{tabular}
\label{t:classes}
\end{center}
\end{table}

\section{Population bins}
\label{app:bins}

\begin{figure*}[htp]
  \label{fig:Population}
    \includegraphics[trim = 20mm 0mm 20mm 0mm, clip, angle=0, width=0.99\textwidth]{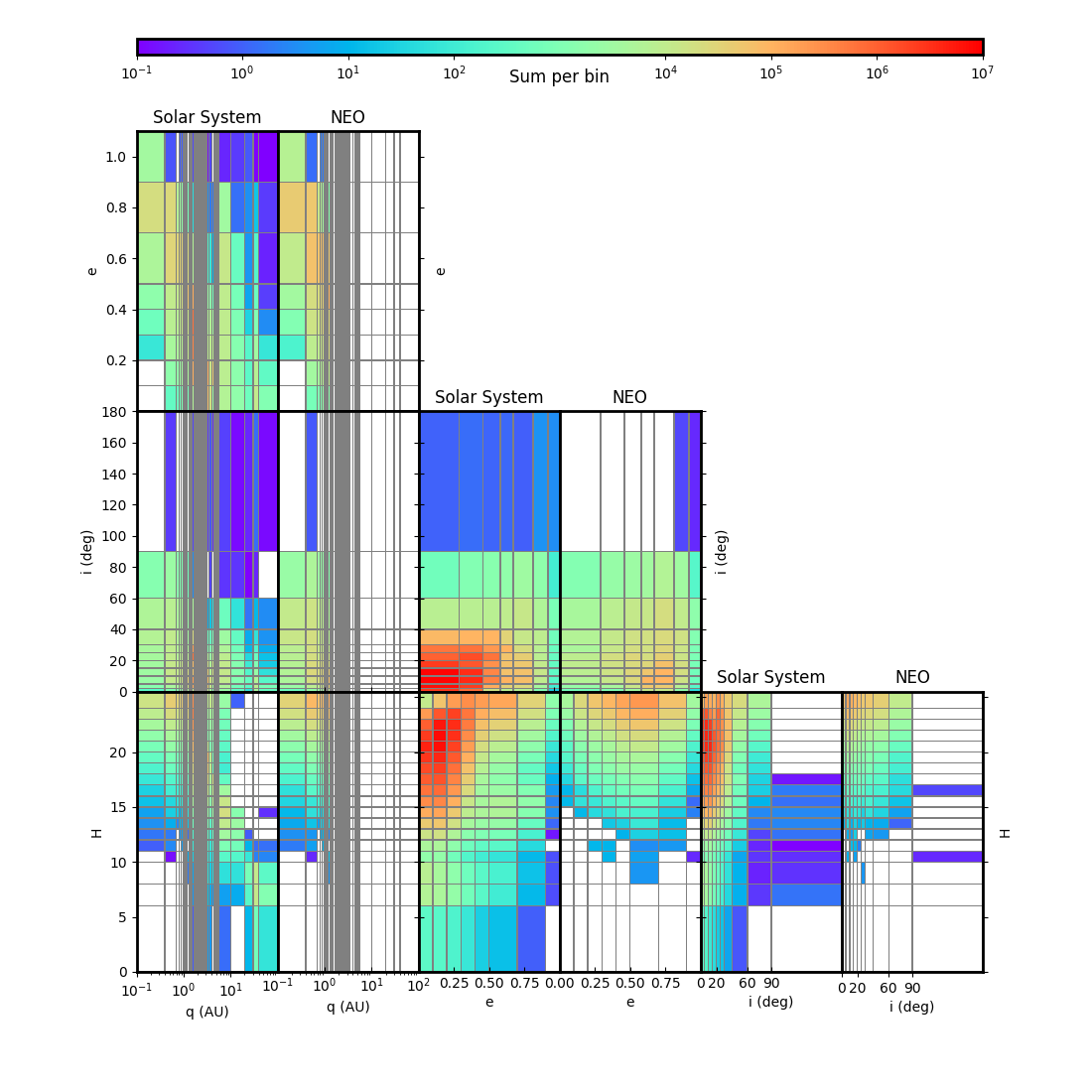}
\caption{Binned population model for the Solar System (odd columns) and NEOs (even columns) as functions of perihelion distance ($q$), eccentricity ($e$), inclination ($i$) and absolute magnitude ($H$) for the full (``raw'') population model. Note the unequal bin sizes and limits of the Solar System model. }
\end{figure*}

In Figure \ref{fig:Population} we illustrate the population model used by \D{} (see Section \ref{s:POPN} above for further description). 


\section{Example Tracklet}
\label{app:Hungaria tracklet}
We provide an example of a tracklet taken from a a typical Hungaria-group asteroid.
The data is presented in a standardized 80-column format\footnote{\url{https://minorplanetcenter.net/iau/info/ObsFormat.html}} - Minor Planet Packed Designation (K18B01E), Mode of Observation ("C" for CCD), Time of observations (Year, Month, Day), Right Ascension (Hours, Minutes, Seconds) and Declination (Degrees, Minutes, Seconds), Magnitude, Band ("w"), Catalog Code ("U"), Packed Reference ("$\sim$2VXl") and Observatory Code ("F51").

\begin{verbatim}
     K18B01E* C2018 01 17.42780 08 01 58.347+41 29 48.20         21.4 wU~2VXlF51
     K18B01E  C2018 01 17.43919 08 01 57.028+41 29 44.58         21.2 wU~2VXlF51
     K18B01E  C2018 01 17.46196 08 01 54.365+41 29 37.55         21.2 wU~2VXlF51
\end{verbatim}


\end{document}